\documentclass[acmsmall]{acmart}

\usepackage{multirow}
\usepackage{amsmath}
\usepackage{subcaption}

\renewcommand\footnotetextcopyrightpermission[1]{} 

\graphicspath{{figures/}}

\AtBeginDocument{%
  }

\usepackage{makecell}

\begin{document}

\title{A Survey of Physics-Informed AI for Complex Urban Systems}



\author{En Xu$^{*}$, Huandong~Wang$^{*}$, Yunke~Zhang, Sibo Li, Yinzhou Tang, Zhilun Zhou, Yuming Lin, Yuan Yuan, Xiaochen Fan, Jingtao Ding, Yong Li}
\email{liyong07@tsinghua.edu.cn}
\thanks{*These authors contributed equally.}
\affiliation{%
  \institution{Department of Electronic Engineering, Beijing National Research Center for Information Science and Technology (BNRist), Tsinghua University}
  \country{China}
  }

\renewcommand{\shortauthors}{Xu et al.}

\begin{abstract}
Urban systems are typical examples of complex systems, where the integration of physics-based modeling with artificial intelligence (AI) presents a promising paradigm for enhancing predictive accuracy, interpretability, and decision-making. In this context, AI excels at capturing complex, nonlinear relationships, while physics-based models ensure consistency with real-world laws and provide interpretable insights.
We provide a comprehensive review of physics-informed AI methods in urban applications. The proposed taxonomy categorizes existing approaches into three paradigms—Physics-Integrated AI, Physics-AI Hybrid Ensemble, and AI-Integrated Physics—and further details seven representative methods. This classification clarifies the varying degrees and directions of physics-AI integration, guiding the selection and development of appropriate methods based on application needs and data availability. We systematically examine their applications across eight key urban domains: energy, environment, economy, transportation, information, public services, emergency management, and the urban system as a whole. Our analysis highlights how these methodologies leverage physical laws and data-driven models to address urban challenges, enhancing system reliability, efficiency, and adaptability. By synthesizing existing methodologies and their urban applications, we identify critical gaps and outline future research directions, paving the way toward next-generation intelligent urban system modeling.
\end{abstract}





\maketitle

\section{Introduction}
The integration of physics and AI is driving a paradigm shift in scientific research. The awarding of the 2024 Nobel Prizes in Physics \cite{nobel2024physics} and Chemistry \cite{nobel2024chemistry} to AI-related achievements and scientists highlights this trend. This not only reflects the far-reaching impact of AI methodologies but also underscores the significant potential of combining physics with data science in addressing complex system problems. Physics provides a foundational framework for explaining the world through mathematical rigor and theoretical elegance, while AI has become a vital tool in modern science due to its ability to capture data patterns and support predictive modeling. These two paradigms excel in different domains. 
For example, Newtonian mechanics formulates natural laws through precise mathematical expressions \cite{newton1850newton}, whereas fields such as protein structure prediction \cite{nobel2024chemistry} and weather forecasting \cite{bi2023accurate} rely heavily on efficient data-driven modeling \cite{he2016deep}.

Urban systems involve multidimensional complexity and dynamic interactions, posing challenges that are difficult to address through traditional physics-based or purely data-driven approaches alone \cite{ding2024artificial,zhang2025metacity}. In particular, for tasks involving nonlinear relationships, such as traffic flow or population migration, AI methods have demonstrated strong capabilities in extracting complex patterns from large-scale historical data and achieving accurate predictions. AI is also effective in handling large-scale, heterogeneous datasets, and is widely applied in domains such as intelligent transportation, power grids, and emergency response, where it supports real-time decision-making and enhances system responsiveness \cite{lecun2015deep}.
\begin{figure*}[t]
	\centering
	\includegraphics[width=\textwidth]{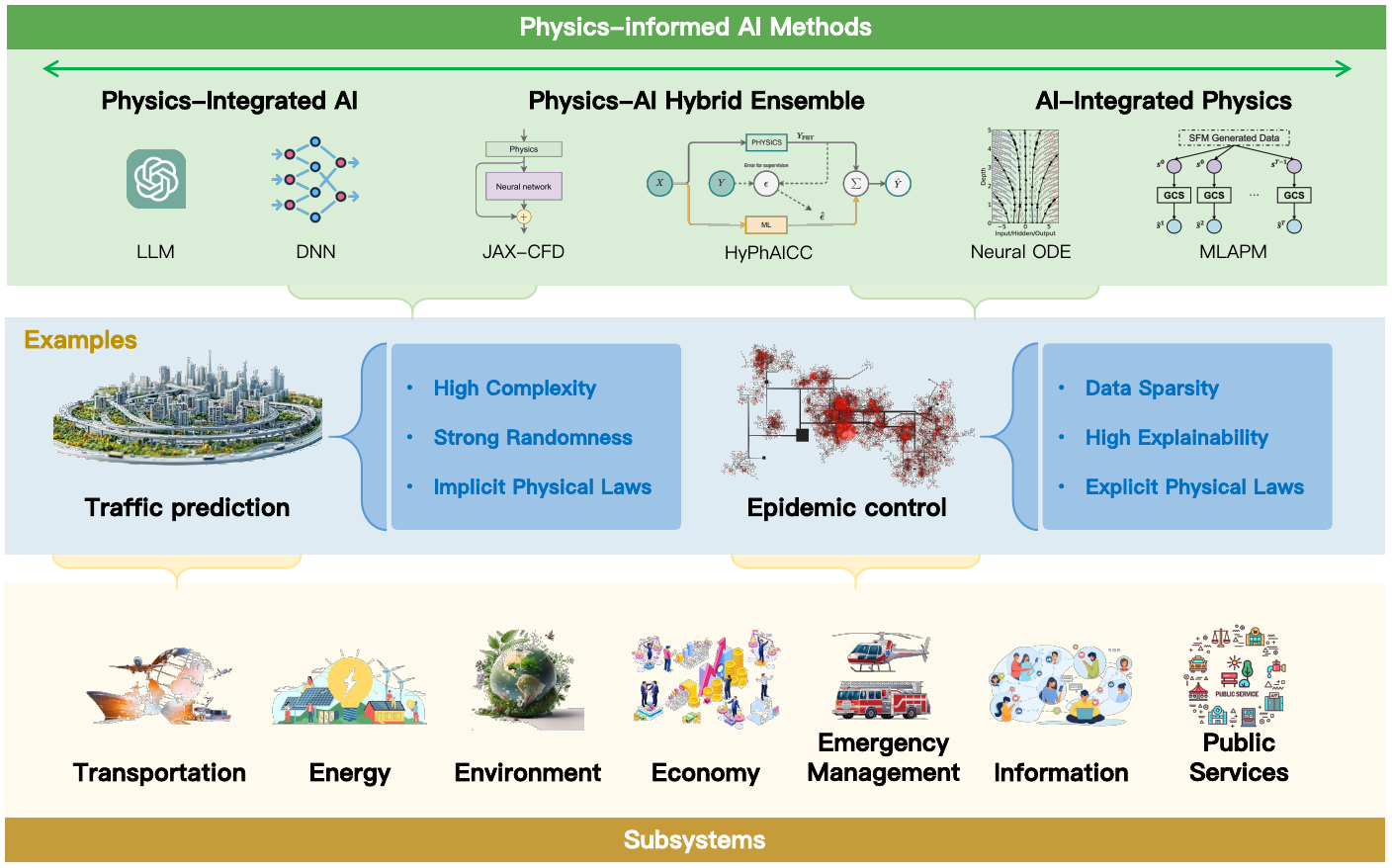}
	\caption{Classification of PIAI methods and their applications in urban systems. The taxonomy categorizes methods into three paradigms—Physics-Integrated AI, Physics-AI Hybrid Ensemble, and AI-Integrated Physics—and reviews their applications across urban subsystems.}
	\label{fig:intro}
\end{figure*}

Many urban problems are fundamentally governed by physical mechanisms, such as fluid dynamics \cite{vinuesa2022enhancing}, heat transfer \cite{shamsaei2022review}, pollutant dispersion \cite{muller2020pollution}, gravity-driven flows \cite{zipf1946p,simini2021deep}, and social force dynamics \cite{helbing1995social}. In such contexts, physics-based models play a critical role in ensuring the reliability and consistency of model outputs, offering particular advantages in multi-scale, dynamic systems and in data-scarce scenarios where prior knowledge mitigates the need for large datasets. To address the complex challenges of urban systems, integrating physics with AI provides a complementary solution: physics offers a rigorous theoretical foundation, while AI leverages data to uncover latent patterns and compensate for the limitations of purely physics-based models. For example, in intelligent transportation systems, AI can optimize traffic signal control while physics-based models maintain the physical consistency of traffic flow; in pollution forecasting, combining meteorological models with AI techniques enhances predictive accuracy. Such physics-AI fusion supports more comprehensive and reliable decision-making, with applications spanning climate adaptation, resource management, and beyond.

The integration of physics and AI varies across urban tasks. Traffic prediction, driven by abundant sensor data and the stochasticity of human behavior and multimodal transportation, typically favors data-driven approaches. Conversely, epidemic control relies more heavily on explicit physical modeling of disease dynamics (e.g., SIS, SIR models \cite{dhar2021genomic}). As illustrated in Fig.~\ref{fig:intro}, we categorize physics-informed AI methods into three paradigms—Physics-Integrated AI, Physics-AI Hybrid Ensemble, and AI-Integrated Physics—reflecting distinct integration strategies. 

Most existing reviews focus on technical classifications of physics-informed AI methods, typically from an AI-centric perspective, emphasizing algorithmic differences. However, these studies generally lack a systematic classification based on the direction and degree of integration between physics and AI, and rarely provide a focused review on urban systems. As shown in Table~\ref{tab:surveys}, this review introduces a new taxonomy—categorizing methods into three paradigms based on their integration strategy—and applies this framework to systematically examine the development of physics-informed AI methods in urban systems.

Furthermore, we divide urban systems into seven core subsystems, ranging from domains of energy and environment to upper-layer services, including public service and emergency management. We systematically examine the interplay between physical mechanisms and data characteristics within each subsystem and explore corresponding modeling requirements for fusion. The structure of this review is as follows: Chapter 2 introduces the overarching framework for physics-informed AI methods and details seven representative fusion methods; Chapter 3 presents the seven key urban subsystems and their underlying physical principles, highlighting the necessity of incorporating physical knowledge into complex urban modeling; Chapter 4 offers a subsystem-by-subsystem review, discussing major research challenges and summarizing practical implementations of existing fusion methods; Chapter 5 concludes the review with a discussion and outlook on future research directions; Chapter 6 concludes the review.

\section{Physics-informed AI Methods}
\begin{figure*}[t]
	\centering
	\includegraphics[width=0.95\textwidth]{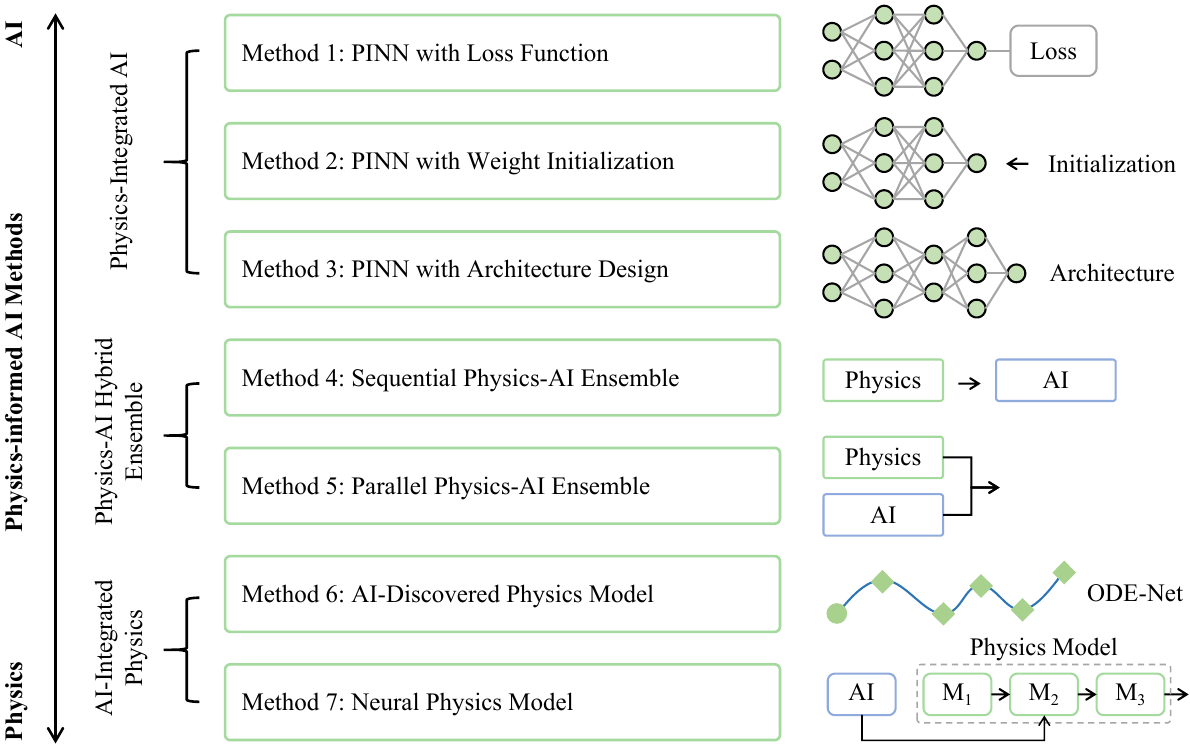}
	\caption{Physics-informed AI methods are categorized into seven types based on the degree of reliance on AI and physics.}
	\label{fig:methods}
\end{figure*}
The integration of AI with physics-based modeling has emerged as a pivotal research direction for tackling complex system challenges. Traditional physics-based models, grounded in explicit mathematical formulations and fundamental laws, provide physically consistent predictions but often face computational inefficiencies and challenges in handling multi-scale dynamics. In contrast, AI-driven methods excel at capturing complex, nonlinear patterns from large-scale data but lack inherent physical constraints, limiting their reliability in data-scarce or extrapolation scenarios.

To bridge these complementary strengths, physics-informed AI (PIAI) methods integrate physics-based knowledge with AI techniques to enhance predictive accuracy, generalization, and computational efficiency. We categorize PIAI methods into three paradigms—(1) Physics-Integrated AI, (2) Physics-AI Hybrid Ensemble, and (3) AI-Integrated Physics—each representing distinct integration strategies. Specifically, Physics-Integrated AI includes Physics-Informed Neural Networks (PINN) with Loss Function, Weight Initialization, and Architecture Design; Physics-AI Hybrid Ensemble comprises Sequential Physics-AI Ensemble and Parallel Physics-AI Ensemble; and AI-Integrated Physics covers AI-Discovered Physics Model and Neural Physics Model. As illustrated in Fig.~\ref{fig:methods}, these methodological approaches enable the synergistic integration of physics and AI to advance predictive modeling and decision support in complex urban systems.

\subsection{Method 1: PINN with Loss Function}
\begin{figure*}[t]
	\centering
	\includegraphics[width=0.95\textwidth]{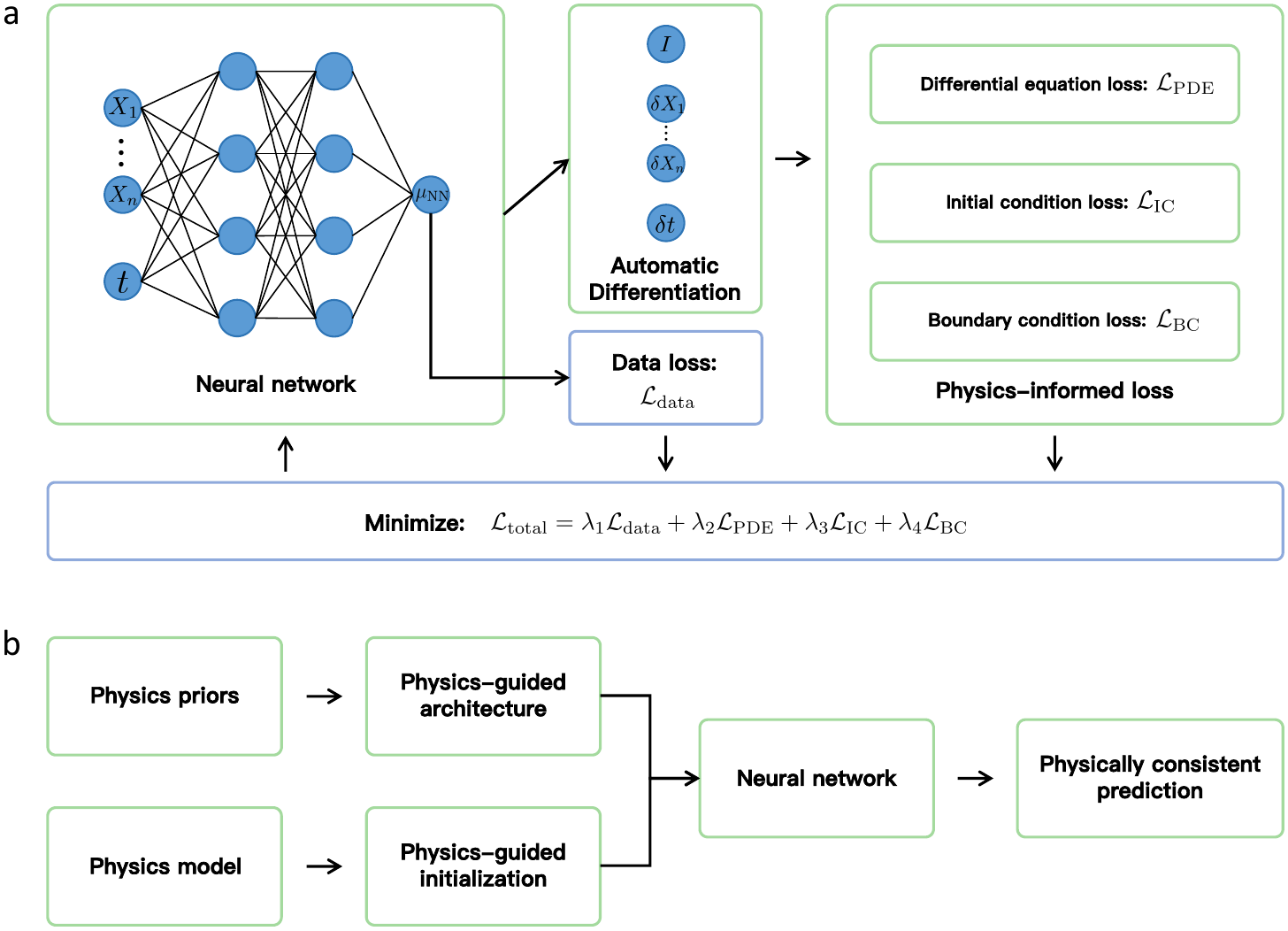}
	\caption{The figure illustrates three AI-dominated PIAI methods: (a) PINN with Loss Function, where a neural network minimizes a total loss combining data and physics-informed losses, and (b) PINN with Initialization and Architecture, which integrates physics priors and models to enhance network design and ensure physically consistent predictions.}
	\label{fig:method1+2}
\end{figure*}
PINN with loss function design integrates physical constraints into deep learning models by encoding fundamental laws as additional loss terms, as illustrated in Fig. \ref{fig:method1+2}a \cite{muralidhar2020phynet}. This helps ensure physically consistent predictions throughout optimization, improving model stability, generalization, and extrapolation, especially in cases with limited or noisy data.

As early as the 1990s, researchers began exploring the use of simple neural networks to approximate solutions of PDEs, with pioneering work by Dissanayake and Lagaris introducing the integration of neural networks with boundary conditions \cite{dissanayake1994neural}. In the 2000s, advances in computational power and tools such as automatic differentiation enabled the application of deeper and more complex neural networks to PDE problems \cite{ozbay2021poisson}. In 2017, Raissi et al. \cite{raissi2019physics} systematically proposed the PINN framework, embedding PDE residuals into the neural network loss function, thereby enabling the simultaneous treatment of forward and inverse problems in an unsupervised setting—marking the formal emergence of modern PINNs. Since then, a variety of variants have rapidly emerged, including conservative PINNs (CPINNs) \cite{jagtap2020conservative}, variational hp-PINNs \cite{kharazmi2021hp}, and physics-constrained neural networks (PCNNs) \cite{liu2021dual}, designed to address diverse modeling requirements in physical systems. In recent years, theoretical studies on the generalization, convergence, and error analysis of PINNs have gradually progressed, laying a preliminary mathematical foundation for their role as a tool in scientific machine learning.

\subsection{Method 2: PINN with Weight Initialization}
PINN with weight initialization leverages prior physical knowledge to optimize the initial state of neural networks, thereby improving training efficiency, stability, and generalization, as illustrated in Fig. \ref{fig:method1+2}b \cite{jia2021physicsSDM}. By embedding physical consistency into the initialization process, models can exhibit better convergence properties and reduce reliance on extensive real-world data.

Early studies demonstrated that pretraining with synthetic data generated by physics-based models can significantly enhance model robustness and generalization to unseen scenarios \cite{jia2021physics}. This approach has since evolved to include techniques such as transfer learning, self-supervised pretraining \cite{jia2021physicsSDM}, and physics-informed Gaussian processes \cite{yan2011bayesian}. These methods enable neural networks to incorporate domain-specific physical knowledge from the outset of training, accelerating convergence and improving predictive accuracy in data-scarce environments.

\subsection{Method 3: PINN with Architecture Design}
PINN with architecture design incorporates physical laws and domain knowledge directly into model structures, enhancing both interpretability and physical fidelity, as illustrated in Fig. \ref{fig:method1+2}b \cite{daw2020physics}. By embedding physical constraints into the network architecture, models can better represent dynamical systems and achieve physically consistent predictions.

Initial efforts in this direction focused on embedding intermediate physical variables or fixing structural parameters \cite{daw2020physics}. Subsequent advances have led to architectures that encode physical symmetries \cite{ling2016reynolds}, Hamiltonian structures \cite{greydanus2019hamiltonian}, PDE-based mappings \cite{chen2018neural}, and Fourier operators \cite{li2020fourier}. These designs constrain the solution space to physically plausible regimes, resulting in more efficient training, stronger generalization, and improved interpretability \cite{greydanus2019hamiltonian}. As a result, physics-guided architecture design has become a key paradigm in building trustworthy models for scientific machine learning and complex system modeling.

\subsection{Method 4: Sequential Physics-AI Ensemble}
\begin{figure*}[t]
	\centering
	\includegraphics[width=0.98\textwidth]{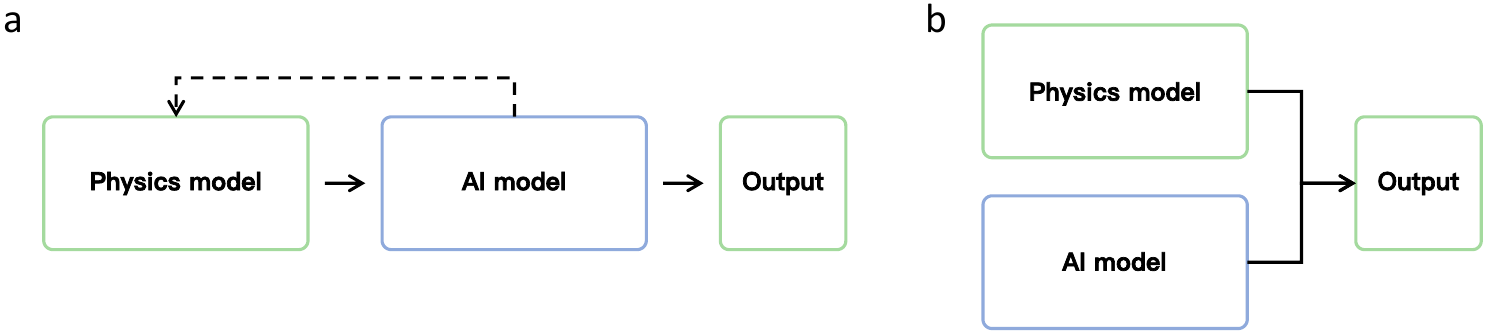}
	\caption{The figure illustrates two integration strategies: (a) pipeline integration, where models are connected in sequence; (b) prediction fusion, where models run in parallel and their outputs are combined.}
	\label{fig:method3+4}
\end{figure*}
Sequential physics-AI ensemble integrates the strengths of both approaches through staged or alternating processes, ensuring physical consistency while leveraging the flexibility of data-driven learning, as illustrated in Fig. \ref{fig:method3+4}a \cite{kochkov2021machine}. This method is particularly effective for complex dynamical systems and computationally intensive simulations, where physics-based models provide structured representations, and AI compensates for unmodeled nonlinearities.

The sequential hybridization of AI and physics-based models has evolved progressively from error correction to predictive enhancement. The earliest form, known as residual modeling, employs machine learning to predict the systematic bias of physics-based models, which is then added as a correction term to the original output \cite{san2018machine}. This was followed by more general pipeline architectures, in which the output of a physics-based model is used as input to a machine learning model, enabling data-driven components to further improve predictive accuracy and generalization \cite{daw2022physics}. In certain temporal modeling tasks, this approach has been extended to iterative structures at each time step, where the physics-based model provides fundamental constraints and the AI model refines predictions accordingly. The two models operate alternately over time, jointly guiding the system’s evolution, allowing the hybrid model to maintain physical consistency while adapting to complex patterns in the data \cite{vlachas2022multiscale}.

\subsection{Method 5: Parallel Physics-AI Ensemble}
Parallel physics-AI ensemble is a modeling strategy in which both components independently process the same inputs and their outputs are subsequently combined through fusion mechanisms to generate final predictions, as illustrated in Fig. \ref{fig:method3+4}b. Compared to sequential coupling, this structure is more symmetric and flexible, allowing the integration of physical constraints with data-driven corrections, and is widely used in complex systems where data are limited or physics models are incomplete \cite{yao2018tensormol}.

Initially employed to capture complementary behaviors across physical scales, this approach has evolved to incorporate dynamic weighting schemes based on spatial, temporal, or error-based criteria \cite{chen2018pga, vlachas2018data}. In tasks such as PDE solving, hybrid architectures have emerged where neural networks and analytical solvers jointly and independently solve different components of the system, improving both accuracy and stability \cite{malek2006numerical}.
The core principle of parallel hybridization lies in using physics-based models to provide structural and theoretical constraints, while AI models compensate for residual errors and capture fine-grained dynamics. The two components reinforce each other within a unified framework. Representative applications include air quality forecasting that fuses pollutant transport models with deep learning corrections \cite{tian2024air}, biochemical reaction systems where AI refines physical residuals \cite{thompson1994modeling}, and fatigue crack growth modeling where AI adjusts for environmental effects to improve structural health predictions \cite{dourado2020physics}.

\subsection{Method 6: AI-Discovered Physics Model}
\begin{figure*}[t]
	\centering
	\includegraphics[width=0.95\textwidth]{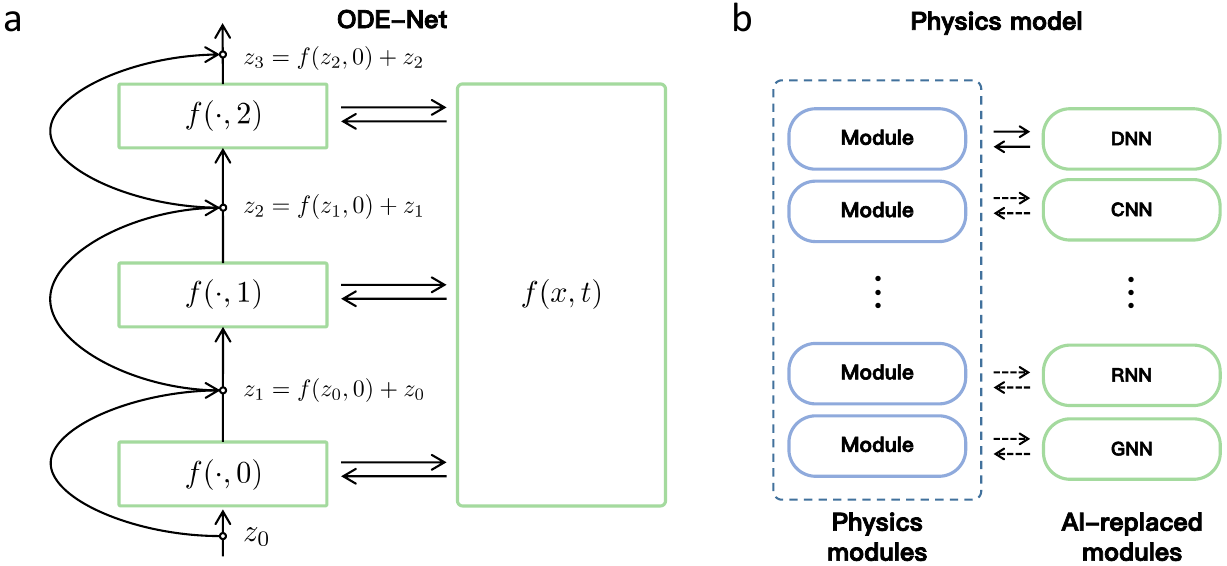}
	\caption{The figure illustrates: (a) discovering governing equations, and (b) replacing parts of a physical model with AI components.}
	\label{fig:method5+6}
\end{figure*}
AI-Discovered Physics Models aim to automatically identify the governing mathematical structures of dynamical systems directly from observational data, or to approximate these structures when the underlying equations are only partially known, as illustrated in Fig.~\ref{fig:method5+6}a. This direction has advanced rapidly in recent years, marking a shift from traditional handcrafted modeling toward interpretable, data-driven scientific discovery.

Early efforts relied on heuristic strategies and expert knowledge to rediscover known empirical laws from synthetic data. For instance, the BACON system used enumeration and pattern matching to recover algebraic relations without differential forms \cite{langley1983rediscovering}. These approaches were later extended to real-world dynamical systems such as ecological modeling, where differential equations are inferred directly from time-series data \cite{bongard2007automated}.
Subsequent developments introduced sparse regression and symbolic modeling as core techniques. The SINDy framework (Sparse Identification of Nonlinear Dynamical Systems) uses sparse selection from a predefined dictionary of nonlinear functions, coupled with numerical derivative estimation, to construct governing differential equations automatically \cite{rudy2017data}. More recently, hybrid methods have combined neural networks with symbolic regression. For example, Udrescu et al. proposed a recursively structured model that first identifies symmetry and separability in the data, then constrains the symbolic search space, improving both discovery efficiency and physical interpretability \cite{udrescu2020ai}.
Overall, this methodology represents a shift from black-box prediction to white-box modeling, offering strong interpretability, generalizability, and theoretical insight—especially in high-dimensional dynamics, multi-source data integration, and scientific hypothesis generation.

\subsection{Method 7: Neural Physics Model}
Neural Physics Models retain the overall structure of physics models while replacing specific physics modules—typically those that are difficult to model, inaccurate, or computationally expensive—with AI-replaced modules. By embedding data-driven components within a physically consistent framework, they enhance nonlinear expressiveness and generalization while preserving interpretability and structural reliability, as illustrated in Fig.~\ref{fig:method5+6}b.

Early applications emerged in computational physics and engineering. Parish et al. replaced turbulence closure terms in RANS models with neural networks to reduce systematic bias in complex flows \cite{parish2016paradigm}, while Hamilton et al. and Zhang et al. used ML to substitute sub-equations or optimization steps in power system estimation, improving accuracy and real-time performance \cite{zhang2019real}.
More recently, this approach has been extended to social and dynamical systems. Examples include replacing interaction terms in pedestrian dynamics models, using neural networks to learn Koopman eigenfunctions without manual observable selection, and employing physics-informed GANs to substitute subgrid closure models in LES, improving efficiency while ensuring physical consistency \cite{zhang2022physics}.

\section{Complex Urban Systems and Involved Physical Laws}\label{sec:physicallaws}
\begin{figure*}[h]
	\centering
	\includegraphics[width=0.9\textwidth]{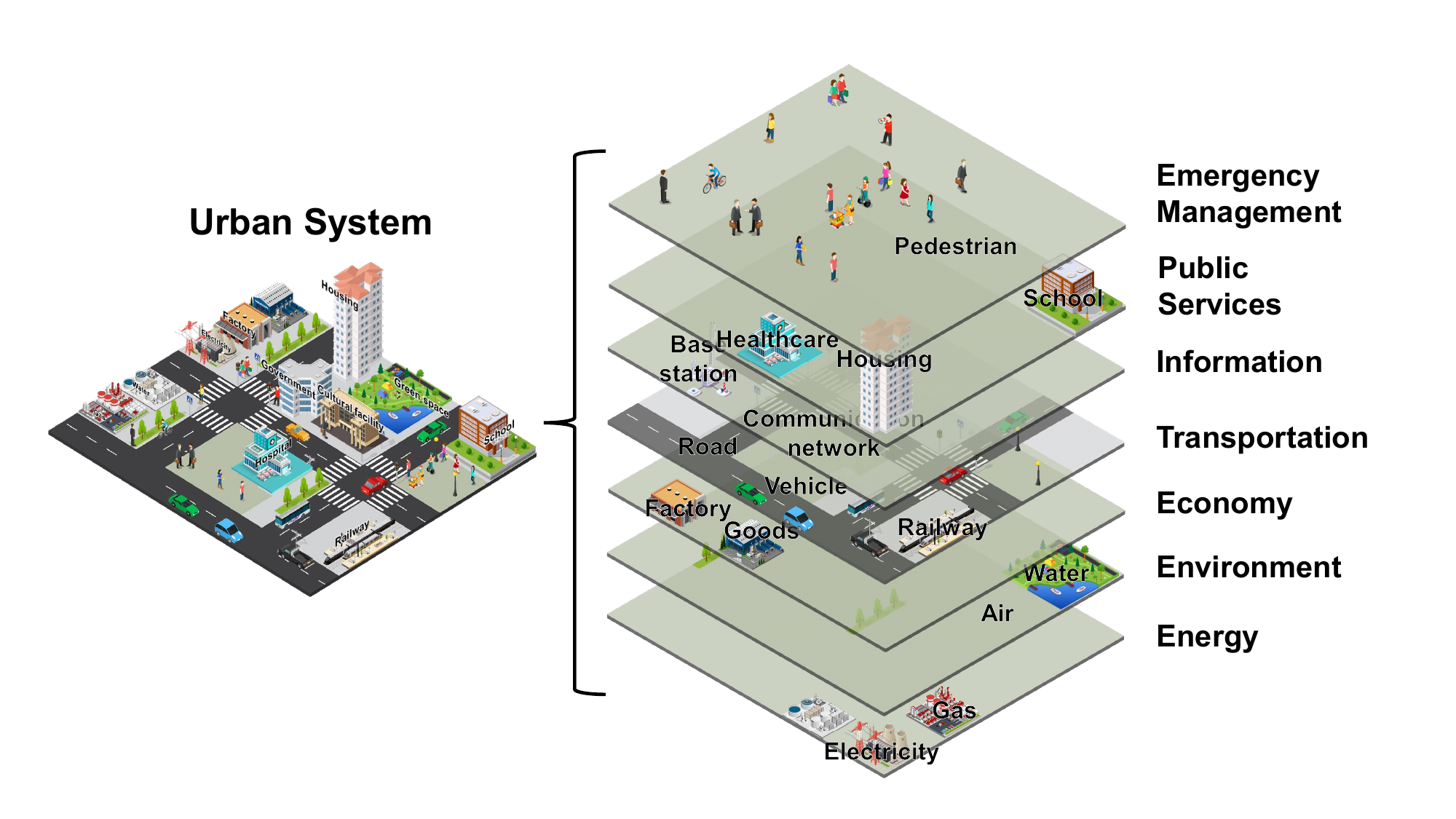}
	\caption{Sub-system of urban systems.}
	\label{fig:urban_system}
\end{figure*}

Urban systems are complex and dynamic, consisting of various subsystems that interact with each other to maintain the function and development of a city. Based on existing studies~\cite{shi2021urban}, in this survey, we broadly category the research areas in urban systems into energy, environment, economy, transportation, information, public service, and emergency management domains, as shown in Fig.~\ref{fig:urban_system}. 
Many elements in these domains are governed by physical laws, as shown in Table~\ref{tbl:physical_laws}. Each physical model represents a crucial aspect of urban dynamics, ranging from power flow in electricity grids to pollution dispersion in the atmosphere, consumer behavior in economic markets, and traffic flow on transportation networks. The study of these physical laws provides insights into the behavior and optimization of urban systems, helping to design more efficient, sustainable, and resilient cities. In this section, we present an overview of each domain and several typical physical models therein.

\subsection{Energy}
The energy system is fundamental to urban operation, encompassing electricity and gas networks that supply power to residential, commercial, and industrial sectors. These systems are governed by physical laws that capture how energy flows, transforms, and balances within a city. For electricity networks, power flow equations describe the relationship between voltage angles, power injections, and network admittance~\cite{smith2022effect}, reflecting conservation laws and the dynamics of electrical interactions. Similarly, gas networks are modeled through pressure balance equations~\cite{zlotnik2015optimal}, which embody fluid dynamics principles to ensure stable distribution across the system. These physical laws—covering energy conservation, resistance, and flow dynamics—form the foundation for managing and optimizing urban energy infrastructure, highlighting the interconnected and law-governed nature of energy systems in cities.
PIAI methods are essential in optimizing power generation, transmission, and consumption, particularly with the integration of renewable energy sources, by enhancing system efficiency and stability \cite{fioretto2020predicting,pombo2022assessing}.

\subsection{Environment}
Urban environmental systems involve complex physical processes governing water flow, air pollution, and noise propagation. These phenomena are described by foundational physical laws that help us understand and manage environmental quality. For instance, water systems follow conservation laws as captured by the Storm Water Management Model (SWMM)~\cite{chang2015novel}, which relates rainfall, inflow, and outflow to changes in water depth across drainage areas. Air quality is governed by diffusion and advection principles, modeled by the Gaussian plume equation~\cite{arystanbekova2004application}, which predicts pollutant dispersion based on emission rates, wind speed, and atmospheric conditions. Noise in urban areas follows the physics of wave propagation, with sound levels decaying logarithmically over distance~\cite{bies2023engineering}. These models illustrate how environmental subsystems are shaped by universal physical laws, providing a scientific basis for pollution control and sustainable urban planning.
In summary, the integration of PIAI methods in urban environmental systems is critical for optimizing air quality, water resource management, soil dynamics, and waste management. These approaches combine physical laws with AI techniques to enhance urban environmental governance and sustainability, providing actionable insights for improving urban ecological systems' resilience \cite{adombi2024causal,he2023novel}.

\subsection{Economy}
Urban economies involve complex interactions between supply, demand, and consumer choice, many of which can be described using physical analogies and laws. The distribution of goods and services is often modeled through network flow frameworks~\cite{ahuja1993network}, reflecting principles of flow conservation across supply chains and trade routes. These models capture how resources move through urban space under constraints similar to those in physical systems. Meanwhile, consumer behavior—central to economic dynamics—is often represented by probabilistic choice models like the Huff model~\cite{huff1963probabilistic}, where decisions are influenced by factors such as distance and attractiveness, resembling potential fields and decay functions in physics. These physical perspectives offer a unifying lens to understand economic behavior in cities, enabling more systematic planning and optimization of urban markets.
PIAI in urban economics addresses key challenges such as optimizing resource allocation, forecasting demand, and improving financial systems. These approaches enhance system intelligence by combining physical laws with AI techniques, leading to more efficient economic decision-making processes \cite{aboussalah2024ai,adombi2024causal}.

\subsection{Transportation}
Transportation systems are essential to urban functionality, enabling the movement of people and goods through roadways, railways, and multimodal networks \cite{rong2024interdisciplinary}. Underpinning these systems are physical laws that describe motion, resistance, and flow dynamics. For railways, mechanical forces such as rolling resistance and aerodynamic drag govern train dynamics and energy consumption~\cite{nash2004railroad}. In road networks, traffic flow is shaped by vehicle interactions and density-dependent dynamics, often modeled through equations that relate speed, acceleration, and spacing~\cite{behrisch2011sumo,treiber2000congested}. These models, grounded in classical mechanics and fluid analogies, reflect how individual behaviors and collective flows emerge from fundamental physical interactions. By leveraging such physical laws, urban planners and engineers can simulate, analyze, and optimize transportation networks for improved efficiency, safety, and resilience.
PIAI in urban transportation addresses key challenges such as traffic flow prediction, state estimation, and real-time control. By combining physical models of traffic dynamics with AI techniques, these methods enhance the efficiency, safety, and adaptability of transportation systems \cite{huang2020physics,tang2024physics}.

\subsection{Information}
The Information domain underpins the digital infrastructure of urban systems, enabling real-time communication and coordination across sectors. Governed by fundamental laws of information theory and signal propagation, this domain ensures the efficient transfer of data essential for urban operations. Models such as the Shannon capacity formula~\cite{shannon1948mathematical} capture the theoretical limits of data transmission based on bandwidth and signal-to-noise ratio, while ray tracing and path loss models~\cite{rappaport2024wireless} describe how signals attenuate over space and interact with the urban environment. These physical principles reveal how information flows are constrained and optimized, forming the basis for reliable, scalable communication networks in smart cities.
PIAI in urban information systems focuses on improving the prediction of information diffusion and optimizing the modeling of opinion evolution. By combining physical laws with machine learning techniques, these approaches enhance the accuracy of social network content predictions and improve decision-making in dynamic environments \cite{tang2023enhancing,tu2022modeling}.

\subsection{Public Services}
Urban public services—including health, safety, and social welfare—are deeply influenced by collective human dynamics, many of which can be captured through physical models. A prominent example is the SIR model~\cite{kermack1927contribution}, which uses differential equations to describe the spread of infectious diseases based on interactions among susceptible, infected, and recovered individuals. This model reflects underlying principles of dynamic systems and population flows, allowing us to quantify how contagion propagates and stabilizes over time. Such physical representations are key to planning effective responses in health crises, allocating resources, and ensuring resilience in the face of emergencies.
PIAI methods play a crucial role in enhancing urban public services, particularly in public health, safety, and infrastructure planning. These approaches integrate physical models with AI to improve epidemic forecasting, predict crowd dynamics, and optimize urban service delivery, thus contributing to the overall resilience and efficiency of urban systems \cite{rodriguez2024machine,zhang2022physics}.

\subsection{Emergency Management}
Emergency management in urban systems addresses the response to and mitigation of disasters, where the movement and interaction of individuals are governed by physical principles of force and motion. A widely used model in this domain is the Social Force Model~\cite{helbing1995social}, which treats pedestrian behavior as driven by virtual forces—toward goals, away from other individuals, and around obstacles—analogous to Newtonian dynamics. These forces encapsulate concepts such as repulsion, attraction, and inertia, allowing for the simulation of crowd behavior under stress. By grounding human movement in physical analogies, such models provide a quantitative framework for optimizing evacuation strategies and improving safety in complex urban environments.
PIAI in urban emergency management focuses on improving the prediction and management of human mobility and disaster dynamics. By integrating physical models with machine learning techniques, these approaches enhance the accuracy of evacuation planning, resource distribution, and disaster mitigation strategies \cite{li2024physics,lee2024predicting}.

\section{Physics-informed AI in Complex Urban Systems}
\subsection{Energy}
\begin{figure}[h]
    \centering
    \includegraphics[width=\linewidth]{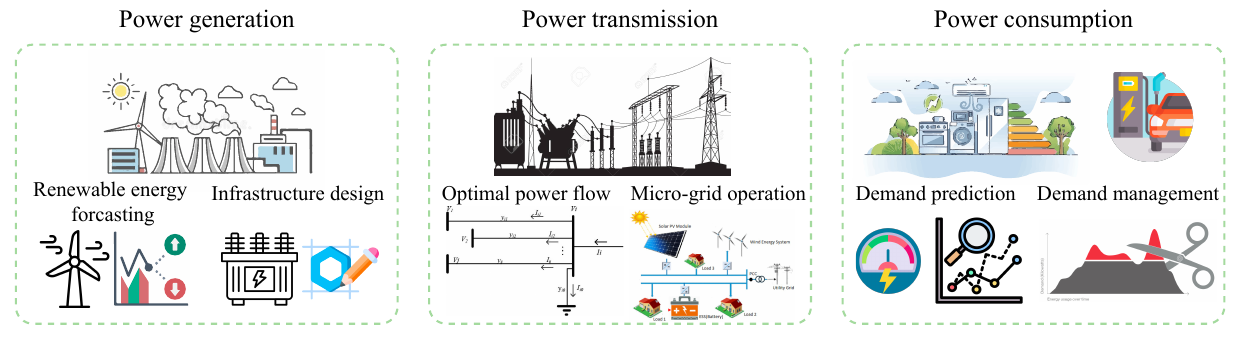}
    \caption{The figure illustrates the PINN applications in urban energy systems, which are further categorized into power generation, transmission, and consumption.}
    \label{fig:4.1}
\end{figure}

\subsubsection{Research Focus and Challenges in Urban Power System} \hfill

\paragraph{Research focus} As renewable energy integration increases, research has focused on improving power system efficiency and reliability. As illustrated in Fig.~\ref{fig:4.1}, we focus on three stages in power system, namely power generation, power transmission, and power consumption. In power generation, key efforts include forecasting renewable energy outputs using weather data and designing advanced power converters. In transmission, the focus is on reducing operational costs while maintaining system stability, such as minimizing voltage fluctuations. In power consumption, better modeling and prediction of energy demand have become crucial to align consumption patterns with the intermittent and variable nature of renewable supply, ensuring more effective grid integration.

\paragraph{Research Challenges} Several challenges hinder power system research. First, ensuring high reliability requires interpretable models with robust worst-case performance, limiting the adoption of complex machine learning techniques like neural networks. Second, large-scale grid operations involve non-linear and non-convex optimization, leading to significant computational costs that restrict real-time applications. Lastly, practical deployment is complicated by limited access to high-quality operational data due to privacy concerns, creating a domain gap between model training and real-world implementation.

\subsubsection{Physical laws in Urban Power System} 
The laws of electrical circuits are the most fundamental laws in power systems, such as Kirchhoff's laws and Ohm's laws. These foundational principles lead to more advanced concepts, including power flow analysis and power system simulation, which are critical for understanding and optimizing the operation of modern grids. In addition, characteristics of specific applications introduce more constraints, such as the temporal power demand patterns of electric vehicles and the operational dynamics of water pumps.

With the integration of renewable energy sources into power systems, new physical considerations have become increasingly important. As shown in Table~\ref{tab:energy}, fluid mechanics plays a pivotal role in analyzing the aerodynamic performance of wind turbines, while solid-state physics is essential for optimizing the efficiency of solar panels. These advancements not only add complexity to power system modeling but also introduce new challenges and opportunities for the application of PINN in power systems research.

\paragraph{Power Generation}  
PIML enhances renewable power generation tasks such as infrastructure design, wind field reconstruction, and generation forecasting by embedding physical principles into machine learning models. For example, co-located wind-solar generation prediction benefits from transforming environmental factors (e.g., temperature, humidity) into physically meaningful coefficients like radiation intensity~\cite{pombo2022assessing}, improving model accuracy. To address sparse wind field observations, 3-D Navier–Stokes equations are integrated into model loss functions, enabling physically consistent spatiotemporal wind field reconstruction~\cite{zhang2021three}. In power electronics, PIML guides automated power converter design through a physics-informed surrogate model with a hierarchical architecture, achieving high accuracy with limited training data~\cite{liu2024physics}.

\paragraph{Power Transmission}  
Power transmission focuses on optimizing energy flow from generators to consumers while maintaining grid stability. A key challenge is the non-convex AC-OPF problem, where a neural network approach incorporating Kirchhoff’s and Ohm’s laws achieves significant computational acceleration—up to $10^4$ times faster—via dual Lagrangian relaxation~\cite{fioretto2020predicting}, with further validation demonstrating robustness under worst-case scenarios~\cite{nellikkath2022physics}. Beyond OPF, PINN-based multi-agent reinforcement learning (MARL) improves voltage control by introducing auxiliary voltage prediction tasks~\cite{zhang2024physics}, while federated MARL with physics-informed rewards promotes local energy utilization and enhances micro-grid self-sufficiency~\cite{li2023federated}.

\paragraph{Power Consumption}
In the power consumption stage, accurate demand prediction and coordination are essential for reducing grid operation costs, especially with the growing reliance on renewable energy. Aligning fluctuating demand with variable power generation requires incorporating physical laws that govern specific consumption patterns.

For instance, PINN improves city-wide electric vehicle (EV) charging demand forecasting by integrating key prior knowledge, such as the geographical proximity of charging stations and the inverse relationship between price and demand~\cite{qu2024physics}. A pre-training strategy is used, where the model first learns from synthetic data based on these principles before adapting to real-world data, enhancing prediction accuracy and capturing complex price-demand dynamics.
In building energy management, PINN leverages the 2R2C thermodynamic model to predict heating energy consumption~\cite{gokhale2022physics}. By embedding this model into the loss function, PINN achieves high accuracy in forecasting temperature and air-conditioning power, particularly for long-term predictions.
For coordination tasks, PINN aids in optimizing urban water pump operations~\cite{ma2024pump}. A multi-agent reinforcement learning (MARL) framework, trained with a physics-informed surrogate model, predicts pipeline pressure and system-wide energy costs. The Navier-Stokes equations are embedded in the loss function, improving prediction accuracy and enhancing cost efficiency.

\subsection{Environment}
\subsubsection{Research Focus and Challenges in Urban Environmental Systems}

\begin{figure*}[t]
	\centering
	\includegraphics[width=\textwidth]{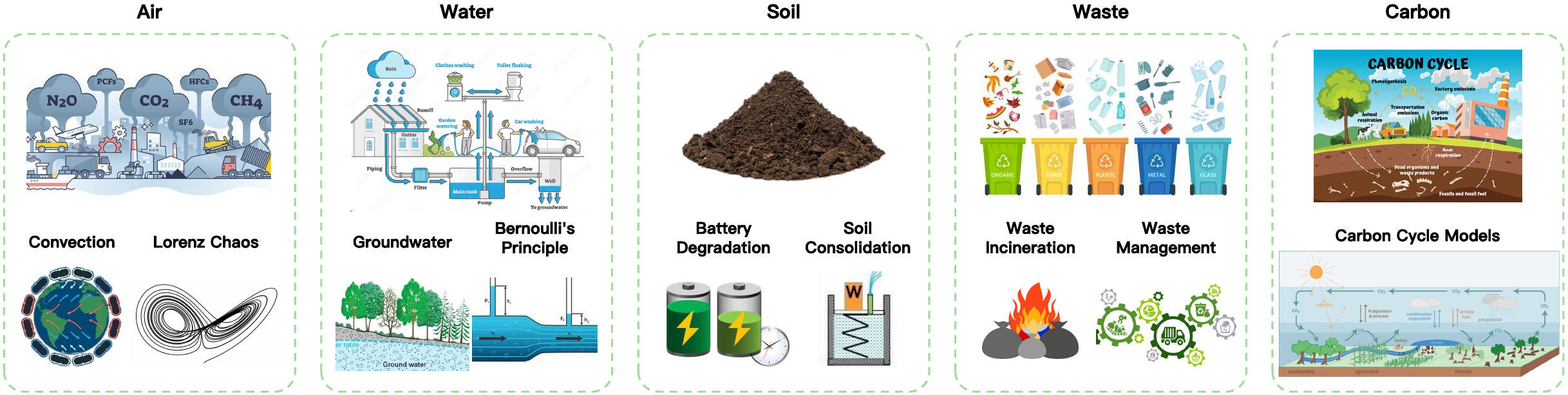}
	\caption{The figure categorizes environmental applications into Air, Water, Soil, Waste, and Carbon, illustrating key processes and their associated physical mechanisms in each domain.}
	\label{fig:environment}
\end{figure*}

Research on urban environmental systems encompasses five key domains: air quality, water resource management, soil dynamics, waste management, and carbon mitigation \cite{airphynet_iclr24,adombi2024causal,lu2022deep,he2023novel,zhang2023physics,bedi2025neural}, as shown in Fig. \ref{fig:environment}. These studies aim to model the physical, chemical, and biological processes shaping urban environments, providing scientific foundations for sustainable development and informed governance.

\paragraph{Research Focus}  
Air quality research focuses on pollutant diffusion, transport, and chemical reactions to guide mitigation. Water management addresses surface runoff, groundwater flow, and urban water cycles for efficiency and flood control. Soil dynamics investigates water transport, thermodynamics, and consolidation under natural and anthropogenic influences, supporting agriculture and disaster prevention. Waste management models degradation and resource recycling to optimize sustainability.

\paragraph{Research Challenges}  
Challenges include expressing complex, nonlinear, and multi-scale physical laws within AI models, addressing data sparsity and noise from limited monitoring, and integrating heterogeneous data effectively. Ensuring physically consistent outputs remains critical, particularly in air and water applications where both accuracy and adherence to physical principles are essential.

\subsubsection{Physical Theories in Urban Environmental Systems}

Urban environmental modeling leverages six categories of physical knowledge—rule-based constraints, physical laws, physical equations, physical models, expert knowledge, and data—as summarized in Table \ref{tab:environment}. Rule-based constraints ensure model rationality, such as pollutant continuity and terrain-aware hydrodynamics \cite{dazzi2024physics}. Physical laws provide deterministic frameworks, including flow conservation, pressure-flow relations, $CO_2$ balance, and heat conduction \cite{ashraf2024physics}. Physical equations, typically expressed as partial differential equations (PDEs), capture pollutant transport, hydrodynamic flows, and soil or waste dynamics \cite{li2023improving,wei2023indoor}. System-level physical models, such as fractional-order Lorenz systems for air dynamics, HBV hydrological models, and Richards’ and Terzaghi’s theories for soil behavior, offer comprehensive representations \cite{liang2019physics,zhang2022physics}. Expert knowledge complements these with domain-specific insights, particularly in waste management \cite{he2023novel}. Additionally, synthetic data generated from physical models mitigates observational limitations and enhances AI training, particularly in data-scarce domains such as water resources \cite{liang2019physics}.

\paragraph{Air}
Physics-AI integration in air quality research combines physical laws with AI to efficiently model and predict complex pollutant transport.
The diffusion-advection equation (DE) was integrated into AirPhyNet, improving prediction accuracy for sparse data and sudden changes \cite{airphynet_iclr24}. The boundary-aware diffusion-advection equation (BA-DAE), combined with neural ODEs, achieved precise multi-scale dynamic predictions \cite{tian2024air}. Finite volume methods (FV) and PDEs, incorporated into PINNs, optimized traffic flow allocation to enhance urban air quality prediction \cite{mei2024traffic}.
The fractional-order Lorenz system, integrated into the SARFIMA-NARX framework, captured nonlinear and chaotic dynamics, reducing RMSE by 20\% \cite{bukhari2022fractional}. Pollutant convection-diffusion PDEs, combined with deep graph neural networks (DGM), enhanced PM2.5 prediction accuracy \cite{li2023improving}. Navier-Stokes equations, integrated with PINNs, improved indoor airflow modeling, reducing errors while enhancing computational efficiency and physical consistency \cite{wei2023indoor}. Finally, a Smart City integrated dashboard systematically combined AI techniques with physical urban models to optimize governance, mobility, and energy strategies, significantly reducing urban air pollution and improving overall sustainability \cite{anwar2024integrating}.

\paragraph{Water}
Physics-AI integration in water resource research combines physical laws and data-driven techniques for efficient hydrological modeling and accurate predictions.
The HBV hydrological model integrated with a causally constrained neural network (H-HBV) improved groundwater level prediction reliability \cite{adombi2024causal}. The Richardson-Richards Equation (RRE), integrated into PINNs, effectively modeled unsaturated soil water flow and solved inverse problems with high efficiency \cite{bandai2022forward}.
A deep neural network (DNN) based on surface runoff dynamics improved runoff and pollution control predictions \cite{liang2019physics}. PI-GNN, incorporating flow conservation and pressure-flow models, optimized large-scale water distribution systems \cite{ashraf2024physics}.
The shallow water equations, integrated into PT-LSTM, enforced mass conservation and enhanced pollutant transport predictions \cite{bertels2023physics}. PI-MR-NN reduced computational costs while improving hydro-mechanical modeling accuracy \cite{zhang2022physics}.
PINNs applied to spherical and enhanced shallow water equations improved flow simulation efficiency \cite{bihlo2022physics}. Finally, the PIDL framework with the Richards equation addressed data scarcity in unsaturated infiltration modeling \cite{dazzi2024physics,lan2024reconstructing}.

\paragraph{Soil}
Physics-AI integration methods in soil research combine physical equations with data-driven techniques to achieve efficient modeling and accurate predictions \cite{lu2022deep,ouyang2024machine}. The partial differential equation (PDE) for two-dimensional soil consolidation, Euler-Bernoulli beam theory, and Richards equation were incorporated into Physics-Informed Neural Networks (PINN), significantly improving computational efficiency and accuracy for complex problems such as consolidation and hydraulic parameter inversion \cite{oikawa2024inverse,zhang2022physics}.
The I2EM model integrated with PIML-FFNN achieved high-precision soil moisture estimation under sparse data conditions \cite{singh2024piml}. The vegetation-soil combustion model enhanced the generalization ability for post-fire soil severity assessments \cite{seydi2024predictive}. The heat transfer equation embedded in PINN with physical constraints improved soil temperature modeling efficiency \cite{xie2024simulating}.

\paragraph{Waste}
Physics-AI integration methods in waste research combine physical knowledge with data-driven techniques to enhance system modeling and prediction capabilities. Domain knowledge of solid waste management systems was integrated into a hybrid neural network (HNN), improving modeling performance and interpretability under data-scarce conditions to support waste management decisions \cite{he2023novel}.
Thermodynamic and dynamic models of battery degradation were incorporated into a physics-informed machine learning framework, enabling non-destructive decoupling of degradation modes and lifecycle prediction. Compared to traditional methods, the model achieved a 25-fold increase in prediction speed and 95.1\% accuracy, significantly reducing validation costs \cite{tao2025non}.

\paragraph{Carbon}
PIAI integration methods in carbon emission research combine physical knowledge with data-driven techniques to enhance modeling and prediction. For example, physics-informed deep networks were used to estimate aboveground carbon biomass (AGB) by integrating physical parameters like solar-induced fluorescence (SIF) and gross primary productivity (GPP) into the network architecture, which improved accuracy, particularly in data-scarce regions \cite{nathaniel2023above}.
Neural Operator models were applied to predict carbon monoxide (CO) concentrations by integrating chemical transport models (CTM) with satellite and ground data. The model enhanced prediction speed and accuracy, especially in extreme pollution events \cite{bedi2025neural}.
These methods improve carbon emission modeling, addressing data imbalance and regional bias, with better spatial resolution and temporal consistency compared to traditional methods \cite{xing2024physically}. However, challenges remain with model robustness and data dependency, which future research should focus on, particularly for global-scale and high-resolution applications \cite{xu2025biogeochemistry}. Real-time early warning systems and multi-source data fusion will be key to further enhancing prediction accuracy \cite{nathaniel2023above}.

\subsection{Transportation}

\subsubsection{Research Focus and Challenges in Urban Transportation}
\begin{figure*}[t]
	\centering
	\includegraphics[width=\textwidth]{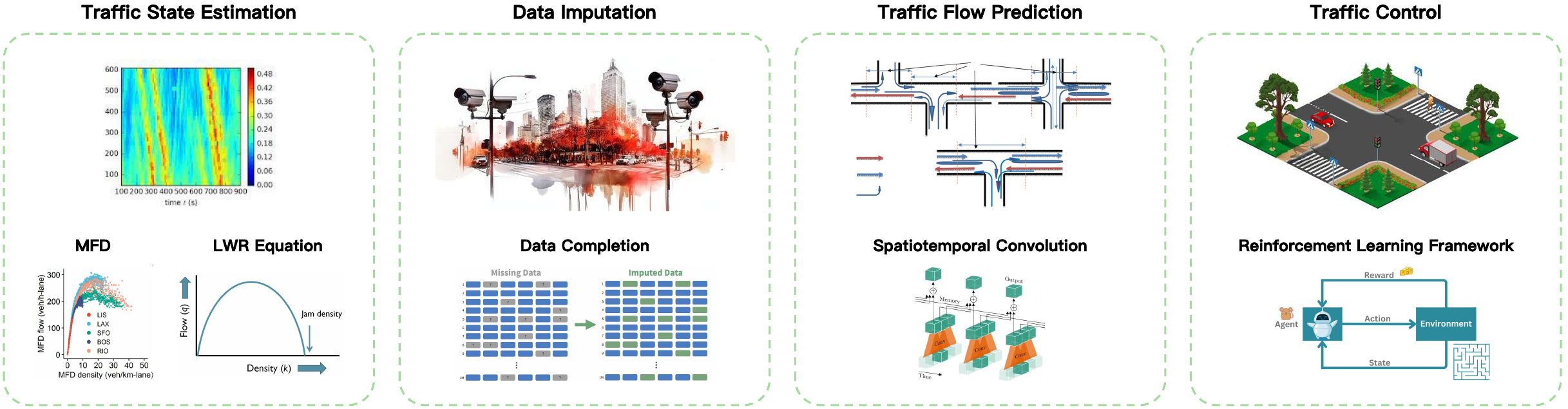}
	\caption{The figure summarizes four core tasks in urban traffic systems—state estimation, data imputation, flow prediction, and control—highlighting key physical and AI-driven modeling approaches in each.}
	\label{fig:transportation}
\end{figure*}

Urban transportation research tackles the complexity of modern traffic systems through improved modeling, prediction, and control \cite{li2023learning,zhang2024noise,zhou2023towards}. As shown in Fig. \ref{fig:transportation}, key tasks include traffic state estimation, data imputation, flow prediction, and control. A growing trend is the integration of physics-based models with AI techniques, where PIAI frameworks embed physical laws into neural networks to overcome the limitations of purely data-driven or model-based approaches.

Key challenges include sparse and noisy data, the complexity of nonlinear, time-varying traffic dynamics, and the difficulty of incorporating physical constraints into flexible AI models. As shown in Table \ref{tab:transportation}, these challenges can be addressed through various strategies that integrate physics with AI. Additionally, ensuring generalization across regions, scenarios, and conditions requires robust domain adaptation. Addressing these challenges calls for interdisciplinary efforts to advance theory and computation, positioning PIAI as a promising direction for adaptive and sustainable urban mobility.

\paragraph{Traffic State Estimation}
Traffic State Estimation (TSE) aims to infer traffic density, flow, and speed from sparse or noisy sensor data, supporting real-time control and long-term planning. Classical models like LWR \cite{lighthill1955kinematic, richards1956shock}, CTM \cite{carlos1994cell}, and ARZ \cite{aw2000resurrection, zhang2002non} describe macroscopic traffic dynamics using the Fundamental Diagram and vehicle conservation.
PIAI provides a data-efficient solution by embedding physical laws into neural networks. Huang et al. \cite{huang2020physics, huang2022physics} integrated LWR and CTM into deep learning models, achieving accurate, robust estimation under limited observations. Observer-based extensions \cite{zhao2022observer, zhao2024observer} further improve performance using only boundary data.
Shi et al. \cite{shi2021physics, shi2021fundamental, shi2021hybrid} proposed learning FD parameters jointly with state estimation, extending to ARZ models to capture complex phenomena like stop-and-go waves. Nonlocal models \cite{zhao2023learning, huang2024incorporating} introduce driver anticipation via integro-differential equations, enabling data-driven learning of anticipation kernels. Ka et al. \cite{ka2024physics} developed the Generalized Bathtub Model to track network-level traffic using time-distance domains.
To address uncertainty, PhysGAN-TSE \cite{mo2022quantifying}, TrafficFlowGAN \cite{mo2022traffic}, and SPIDL \cite{wang2024knowledgedatafusionorientedtraffic} use generative models and autoencoders to quantify and learn uncertainty distributions. For network-scale TSE, PSTGCN \cite{shi2023spatiotemporal} integrates graph convolutions, GRU, and physical constraints, achieving strong performance even with partial data.

\paragraph{Traffic Data Imputation and Calibration}

In real-world traffic applications, missing or unreliable sensor data—caused by hardware failures, environmental disturbances, or human error—significantly hinder accurate traffic state estimation and prediction. Thus, inferring missing variables and calibrating model parameters under such imperfect conditions is a key challenge.
Tang et al.~\cite{tang2024physics} addressed this by integrating the macroscopic CTM model with an unsupervised denoising autoencoder, enabling robust parameter calibration through minimizing the discrepancy between simulated and observed data. Their framework incorporates boundary conditions (e.g., upstream flow) as inputs, supporting conditional generation and handling data noise effectively without requiring labeled data.
Xue et al.~\cite{xue2024network} combined the Network Macroscopic Fundamental Diagram (NMFD) with Graph Neural Networks (GNNs), enforcing physical consistency in interpolation by embedding traffic flow laws, which improves the accuracy and interpretability of estimated traffic states.

\paragraph{Traffic Flow Prediction}
Traffic flow prediction aims to forecast future traffic states using historical and real-time data. While physics-based models offer interpretability and robustness by capturing the underlying traffic dynamics, they often fall short in handling complex patterns. Integrating them with data-driven approaches enhances accuracy, generalization, and computational efficiency.
Yuan et al.~\cite{yuan2022trafficflowmodelingwithgradual} proposed Gradual Physics Regularized Learning (GPRL), which incrementally embeds traffic flow models—from fundamental diagrams to second-order models like PW and ARZ—into Gaussian processes. This hierarchical integration improves prediction accuracy while significantly reducing computational cost.
Ji et al.~\cite{ji2022stden} introduced the Spatio-Temporal Differential Equation Network (STDEN), which models traffic flow as being driven by a latent potential energy field. By embedding this field into neural network architectures, STDEN captures complex dynamics in urban road networks and outperforms existing methods in predictive accuracy.
Li et al.~\cite{li2022network} developed a two-stage physics-informed transfer learning framework for network-level prediction. Their method leverages Macroscopic Fundamental Diagrams (MFDs) to partition networks and applies Deep Tensor Adaptation Networks (DTAN) for domain adaptation between source and target regions. This approach addresses cold-start and distribution shift problems effectively.
Deshpande and Park~\cite{deshpande2024kalman} proposed a Physics-Informed Graph Convolutional Gated Recurrent Neural Network (PI-GRNN) combined with a Kalman Filter mixture model. This architecture handles non-linearity and uncertainty, while physics constraints ensure that learned dependencies reflect true causal relationships.

\paragraph{Traffic Control}
Traffic control focuses on optimizing traffic flow via signals, ramp metering, or other mechanisms, typically formulated as optimization problems. While traditional methods rely on physical modeling, the integration of reinforcement learning (RL) offers greater flexibility and adaptability without sacrificing interpretability.
Su et al.~\cite{su2021adaptive} proposed an adaptive control system that combines kinematic wave-based traffic models with approximate dynamic programming (ADP). By using parametric value function approximators, their decentralized framework reduces computational burden and enables robust control across large-scale, congested networks.
Han et al.~\cite{han2022reinforcement} developed a physics-informed RL framework for ramp metering, leveraging both real-world and synthetic data to improve policy robustness and reliability. Their hybrid approach outperforms classical feedback and pure RL strategies in reducing travel time and alleviating congestion.

\subsection{Economy}
\subsubsection{Research Focus and Challenges in Urban Economics}
\begin{figure*}[t]
	\centering
	\includegraphics[width=\textwidth]{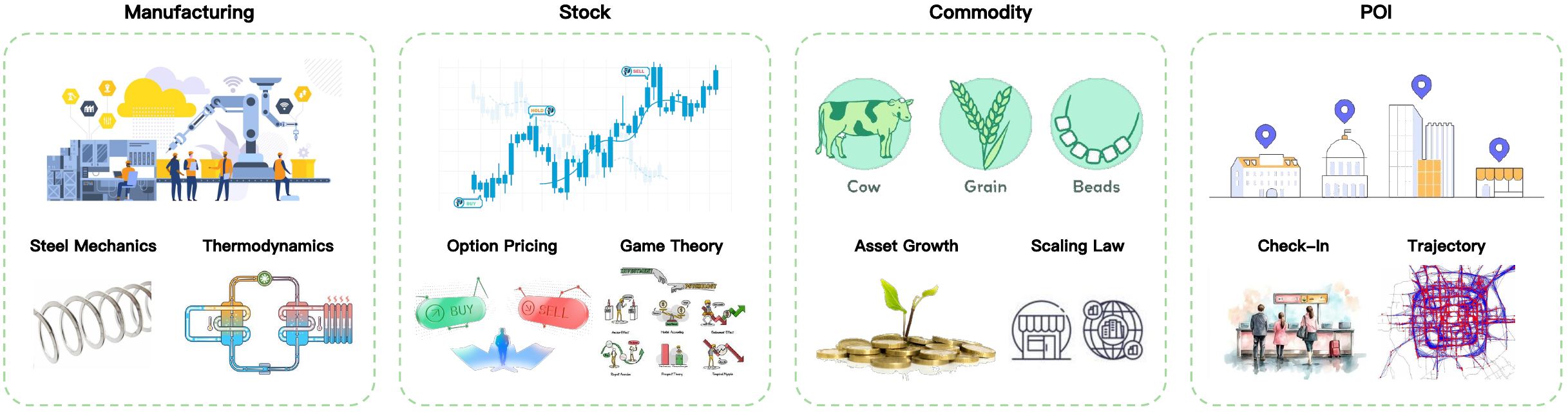}
	\caption{The figure categorizes economic applications into Manufacturing, Stock, Commodity, and POI, illustrating key processes and their associated mechanisms in each domain.}
	\label{fig:economy}
\end{figure*}

Urban economics, a critical domain in urban systems research, encompasses production and manufacturing, stock markets, commodity sales, and point-of-interest (POI) recommendations, as shown in Fig. \ref{fig:economy}. This field aims to address complex dynamics across industrial operations, financial systems, supply chains, and user behavior modeling, emphasizing the integration of physical knowledge and AI for enhanced system optimization and intelligence.

\paragraph{Research Focus}
In production and manufacturing, efforts focus on optimizing resource allocation and energy efficiency in energy-intensive industries, combining physical heat conduction models with AI for real-time prediction. Stock market research addresses nonlinear price dynamics influenced by volatility and arbitrage-free principles, supporting risk management and investment strategies. Commodity sales studies optimize supply chains, forecast demand, and manage inventories by integrating physical models and data-driven methods. POI recommendation predicts user behaviors using historical check-ins, accounting for spatiotemporal dynamics and personalized preferences.

\paragraph{Research Challenges}
Adhering to Physical Laws: Dynamic processes in urban economics must comply with physical principles, such as heat conduction equations and arbitrage-free conditions, yet embedding these into deep learning remains challenging.
Fitting Complex Dynamics: Advanced modeling techniques are required for high-dimensional nonlinear processes, such as stock market volatility, which is described by Heston models.
Data Sparsity: Incomplete datasets, such as missing user trajectories in POI recommendations or limited market data during extreme fluctuations, hinder model performance, necessitating synthetic data generation or imputation.
Customized Deep Learning Architectures: Tailored models, such as embedding dynamic hedging and arbitrage principles in financial applications, enhance compliance with domain-specific physical laws.
Ensuring Stability and Reliability: Models must produce robust outputs under uncertainty, with dynamical constraints improving stability in scenarios like supply chain optimization.

\subsubsection{Physical Theories in Urban Economics}
Urban economics leverages diverse physical theories to model, analyze, and optimize complex systems. These theories span rule-based constraints, physical laws, equations, models, and expert knowledge, playing critical roles in production and manufacturing, stock markets, commodity sales, and POI recommendations, as shown in Table \ref{tab:economy}.

\subparagraph{Rule-based constraints}
, such as symmetry and non-negativity, underpin physical modeling. In production and manufacturing, design standards ensure compliance with safety and operational norms, forming the foundation for system optimization \cite{fu2024physics}.

\subparagraph{Physical laws}
summarize universal principles derived from empirical observations. In manufacturing, laws like voltage and power relationships guide energy optimization. In structural engineering, forces such as axial, shear, and bending moments ensure mechanical stability \cite{wang2022hybrid,fu2024physics}.

\subparagraph{Physical equations}
often expressed as partial differential equations (PDEs), describe system dynamics. In manufacturing, heat transfer PDEs model thermal processes, while in POI recommendation, graph differential equations capture user behavior's spatiotemporal evolution \cite{zobeiry2021physics,yang2024siamese}.

\textit{Physical models}
represent complex systems by capturing their multi-scale dynamics. In manufacturing, the Hottel zone method is used to model heat transfer, while reliability analysis assesses structural safety. In stock markets, the Black-Scholes equation characterizes option pricing dynamics. In commodity sales, scaling laws illustrate company growth, and for POI recommendation, trajectory flow graphs describe user transitions between locations \cite{shi2024physics,tao2024predicting,zhuang2024tau}.

\subparagraph{Expert knowledge}
integrates domain-specific observations to enhance physical modeling. In manufacturing, thermodynamics and material behavior insights drive advanced material design and optimization \cite{raabe2023accelerating}.

\paragraph{Manufacturing}
PIAI has revolutionized production and manufacturing by integrating physical theories to address complex challenges efficiently and accurately.
Heat transfer problems are solved using PINNs, achieving high-accuracy predictions under uncertain boundary conditions while outperforming finite element methods \cite{zobeiry2021physics}. FrameRL combines finite element analysis with reinforcement learning, automating steel structure design with sub-second computation times \cite{fu2024physics}.
The material design leverages thermodynamic principles, enabling efficient discovery of high-entropy alloys with limited data \cite{raabe2023accelerating}. Similarly, the Hottel zone method integrated into PINNs improves furnace heat transfer predictions under sparse data \cite{dutta2024application}.
A hybrid physics-data model (HPDM) enhances interpretability and responsiveness in intelligent manufacturing, reducing costs and increasing automation \cite{wang2022hybrid}. Structural reliability is advanced with PINNC, achieving superior accuracy in nonlinear problems \cite{shi2024physics}.
Finally, the PITA model improves industrial load monitoring, enhancing energy decomposition and efficiency management \cite{huang2022physics}. These innovations demonstrate the transformative impact of PIAI on industrial processes.

\paragraph{Stock}
PIAI has significantly advanced option pricing and risk management. A PINN framework integrates the Black-Scholes equation and free boundary conditions, solving high-dimensional, path-dependent problems efficiently and accurately \cite{gatta2023meshless}. The FINN framework combines no-arbitrage principles and delta-gamma hedging with neural networks, enhancing computational efficiency and robustness \cite{aboussalah2024ai}. These approaches provide precise, scalable solutions for complex financial challenges.

\paragraph{Commodity}
PIAI enhances growth prediction in commodity sales by integrating company growth models (scaling laws and asset dynamics) with LSTM time-series models. This approach captures long-term trends and short-term fluctuations, significantly improving forecasting accuracy and interpretability, and offering a robust solution for company growth prediction \cite{tao2024predicting}.

\paragraph{POI}
PIAI improves Next-POI recommendation by addressing data sparsity and behavior modeling challenges. One approach integrates graph differential equations with Siamese networks, embedding user interest dynamics into time-series graphs to enhance accuracy in sparse, long-span scenarios \cite{yang2024siamese}. Another combines trajectory flow graphs with self-attention and uncertainty modeling, leveraging missing POI insertion and confidence calibration to boost recommendation accuracy and robustness \cite{zhuang2024tau}.

\subsection{Information}
\subsubsection{Reaserch Focus and Challenges in Urban Information Systems}
\begin{figure}
    \centering
    \includegraphics[width=\linewidth]{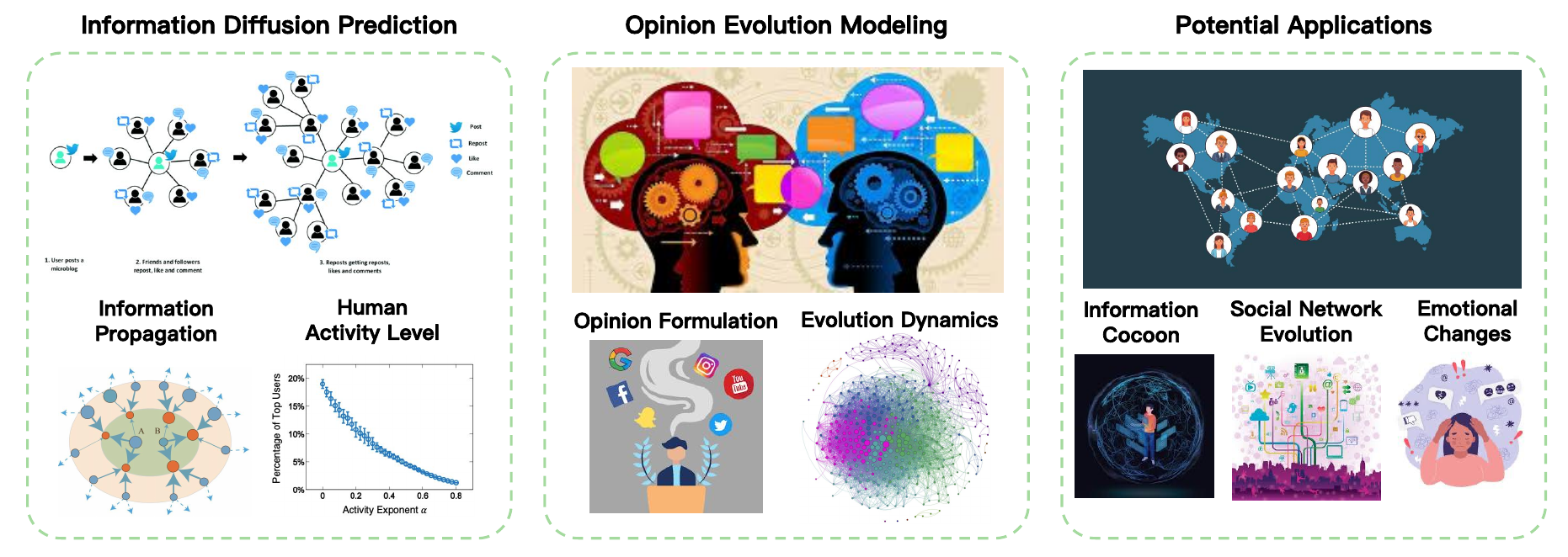}
    \caption{The applications of PIAI in Information Systems.}
    \label{fig:4.5}
\end{figure}

With the development of digitalization, social networks have become an important part of our daily lives. Information in social networks refers to the various forms of data generated through platforms, which often reflect the structure and dynamics of the social network \cite{yuan2023learning}. Modeling the dynamics of information on social networks provides insight into information dissemination mechanisms, such as how messages spread across networks and what factors influence the prevalence of information. Secondly, user experience can be optimized by analyzing user behavior patterns. For example, a personalized recommendation system can provide more accurate content and services according to users' interests. Moreover, social network modeling can help identify community structures and social influencers, which is crucial for marketing strategy development. An illustration of this part is shown in Fig.~\ref{fig:4.5}.

In recent years, with the rapid development of physic-informed machine learning (PIML), researchers have begun to explore new ways to integrate it with the modeling of information dissemination on social networks. This combination not only inherits the accuracy of traditional physics models but also makes full use of the powerful data processing power of machine learning, providing a new perspective for understanding the complex dynamics of information on social networks. In the following part, we will explore the applications of physic-informed AI in information diffusion (also referred to as information propagation and information cascade) prediction and opinion evolution. Then, we will demonstrate the potential applications of physic-informed AI in other scenarios in the information system such as predicting information cocoons and filtering bubbles.

\subsubsection{Physic-informed AI in Information Diffusion Prediction}
Information diffusion is the process by which information is passed on a social network from one or more sources to other members through the social network platform. 
With the development of deep learning technology, more and more researchers have begun to use data-driven methods to predict the spread of information on social networks~\cite{qiao2023rotdiff,sankar2020inf,qiu2018deepinf}.
However, existing methods have proven not to model the dynamics of information propagation on social networks well, especially phase transitions. Xie et al.~\cite{xie2021detecting} model the spread mode of information on social networks by using the frequency of tweets posted by users and their degree on social networks to represent the level of user activity. 

Although user activity levels have been shown to be effective in modeling the spread of information on social networks, existing research has not taken this physical law into account~\cite{tang2024msa}. Tang et al.~\cite{tang2024msa} proposed MSA-Net, a multi-scale information diffusion model that can sense user activity level. Specifically, MSA-Net learns three different levels of network representation (micro, meso, and macro) and introduces the concept of user activity level, which measures the difference between individual users by their connectivity and average number of tweets per unit of time. 
Tu et al.~\cite{tu2022modeling} proposed a predictive model of information Diffusion called ODID (Optimized Diffusion Intensity Dynamics). The authors propose a mathematical framework based on the heat transfer equation to model information density flow in social networks, adjusting the information transfer coefficient $\alpha$ to fit actual data.
Cheng et al.~\cite{cheng2024information} proposed a new framework called CasDO to solve the problem of cascaded popularity prediction in social network content diffusion analysis. It uses the probabilistic diffusion model and ODEs to deal with the temporal irregularity of information cascade events and the inherent uncertainty of information diffusion.
Yu et al.~\cite{yu2024information} proposed PiGCN which combines structural features with temporal features and introduces the concept of physical information neural networks (PINN) to capture dynamic changes in information propagation.

\subsubsection{Physics-informed AI in Opinion Evolution Modeling}
Opinion evolution refers to the process by which users' opinions change over time in a social network~\cite{noorazar2020recent,dong2018survey}. In this process, the positive and negative opinions of users interact with each other and together affect the overall opinion dynamics. Studying the evolution of opinion is helpful in understanding the formation and development mechanism of public opinion, and is of great significance for predicting the change in public attitude, evaluating the effect of policies, and coping with social events. Traditional opinion evolution methods typically model opinions dynamically as discrete and homogeneous processes, meaning that they update user opinions at fixed intervals and assume that all users or nodes follow the same evolution rules~\cite{bi2024duan}. However, these assumptions make the modeling of opinion evolution suboptimal. Therefore, we need to introduce physic-informed AI to model the heterogeneity among users and combine physical laws to realize the modeling and prediction of opinion evolution.

De et al.~\cite{de2016learning} proposes SLANT, a probabilistic framework for modeling and predicting opinion dynamics in social networks. It utilizes labeled jump-diffusion stochastic differential equations to represent changes in user opinions over time and allows efficient simulation of models and estimation of parameters from historical event data.
Okawa et al.~\cite{okawa2022predicting} proposed SINN, which combines large-scale social media data with prior scientific knowledge from sociology and social psychology by transforming traditional opinion dynamics models into ordinary differential equations (ODEs) and approximating these ODEs using neural networks.
Li et al.~\cite{li2024hides} proposed a method called HiDeS NODE. It extends the expressive range of NODE and enhances the predictive capability by introducing the interaction between higher-order derivatives and state variables and using state vectors and their higher-order derivatives as supervisory signals in training.
Duan et al.~\cite{bi2024duan} proposed a framework called Bi-Dynamic Graph Ordinary Differential Equation (BDG-ODE) to simulate the dynamic changes of opinions in social networks. It captures the complexity of opinion evolution through two dynamic processes: the evolution of positive and negative opinions. The model includes a dual opinion encoder that handles both positive and negative opinions separately and models the evolution of opinions over time through the bidirectional graph ordinary differential equation, enabling continuous capture of opinion changes. In addition, an opinion synthesis decoder is introduced, which can effectively map the representation evolved from the potential space back to the opinion space.

\subsubsection{Potential Applications of Physic-informed AI in Information Systems}
In addition to the typical research problems mentioned above, there are a large number of problems in the information space that can be solved with the help of physic-informed AI. Such as information cocoon modeling~\cite{piao2023human}, relationship evolution in social networks~\cite{stokman2013evolution}, emotional and psychological state changes~\cite{schachter1964interaction}, etc. Existing methods have used dynamic equations to describe the evolution of relevant system dynamics, but how to use physic-informed AI to enhance the accuracy and efficiency of modeling is still an open question.
\subsection{Public Services}
\begin{figure}
    \centering
    \includegraphics[width=\linewidth]{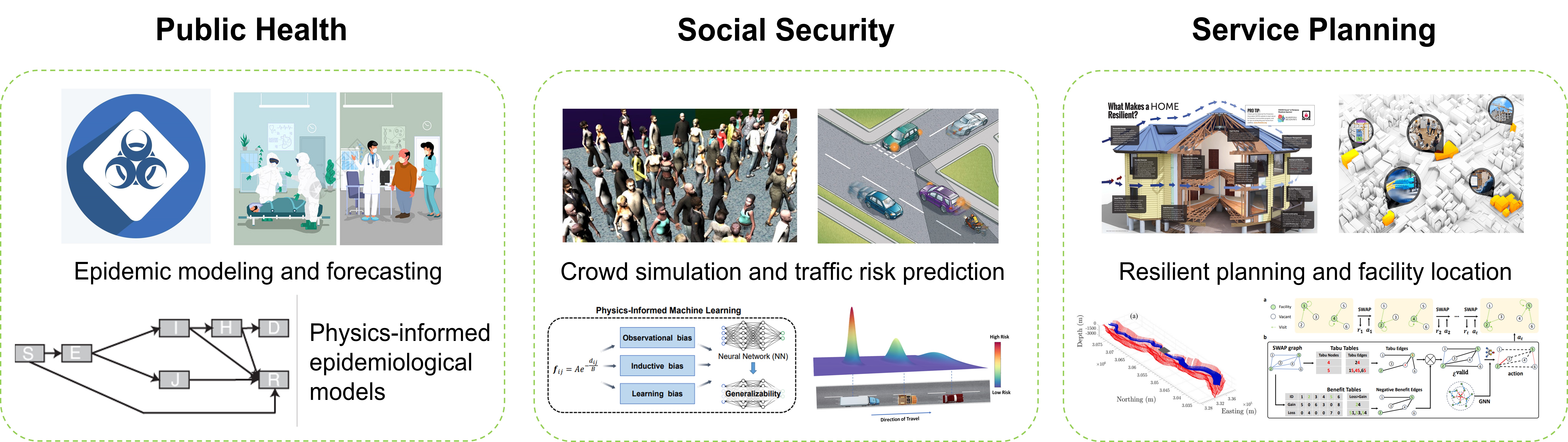}
    \caption{The applications of PIAI in urban public service scenario.}
    \label{fig:4.6}
\end{figure}

\subsubsection{Research Focus in Urban Public Services.}

Urban public services play a crucial role in the urban system by providing essential services that maintain residents' health, safety, and social well-being. \textit{Public health} focus on evaluating and forecasting the overall health status of residents, thus ensuring access to fundamental healthcare and safeguarding the right to health and corresponding subsidies for all citizens. \textit{Safety and social security} aim to accurately predict the movements of urban populations, including crowds and traffics, ensuring public safety and offering support to vulnerable populations. \textit{Service supply and planning} emphasize the provision of cultural, recreational, education, dining, and other community infrastructure, ensuring that urban residents have access to adequate services and facilities to enrich their daily lives. Together, these elements create a cohesive framework that supports the overall functionality and resilience of urban systems. 

\textit{Research Challenges.} Predicting and preventing epidemic outbreaks is a key challenge in urban public health. The COVID-19 pandemic emphasized the need for effective forecasting models, as urban disease spread is influenced by social behavior, mobility, and healthcare infrastructure. Integrating diverse data sources and generalizing across urban contexts complicates this task. Approaches like PINNs and EINNs (Epidemiology-Informed Neural Networks) combine physical principles with data-driven techniques, but issues such as parameter identifiability, uncertainty quantification, and real-time adaptability remain. Urban safety and social security research faces challenges in predicting crowd dynamics and vehicle movements in rapidly urbanizing environments. Traditional models struggle with the heterogeneity of human behavior and the interactions between pedestrians, vehicles, and infrastructure. To address these complexities, advanced approaches like PIAI and machine learning are essential for improving predictive accuracy and real-time decision-making in urban safety management. Urban service supply and planning involve the complex interplay between urban structures, population dynamics, and the accessibility of essential services, necessitating resilient and adaptive solutions. Traditional optimization methods often struggle with the nonlinear and dynamic nature of urban systems, leading to high computational demands and less effective outcomes.

\subsubsection{Physical Theories in Urban Public Services.}

The study of urban public health mainly relies on physical models (often expressed as ODEs) that characterize the spread of epidemics in urban populations with various variants, such as the SIR (Susceptible - Infected - Recovered) model~\cite{beckley2013modeling} and the SEIR (Susceptible - Exposed - Infected - Recovered) model~\cite{he2020seir}. Recent extensions of these models take into account factors such as social behavior, mobility patterns, and functional-order dynamics.

To support the prediction of crowds and road traffic for public safety, it is essential to employ physical theories that can effectively characterize both individual movements and their macro-level patterns. One key physical model is the social force model~\cite{helbing1995social}, which explains how social interactions and pedestrian behaviors are influenced by societal norms, pressures, and the dynamics of group behavior. Another area of research applies physical laws from fields such as fluid dynamics and electromagnetism to accurately characterize individual movements.

Key physical theories in urban service supply and planning include transportation flow, energy exchange in microclimates, and the impact of environmental hazards like floods and earthquakes. These models are crucial for designing resilient infrastructure to withstand urban hazards and climate change.

\paragraph{Public health.} Recent research efforts have demonstrated the power of integrating physics theories with AI techniques to improve epidemic modeling and forecasting. Kharazmi \textit{et al.}~\cite{kharazmi2021identifiability} extend the classical SIR model with PINNs, using neural networks to infer time-dependent parameters and unobserved dynamics, thus enhancing the accuracy of COVID-19 spread predictions across diverse regions. Rodriguez \textit{et al.}~\cite{rodriguez2023einns}  introduce Epidemiology-Informed Neural Networks (EINNs), a novel framework that merges traditional mechanistic models with AI’s capacity to process diverse data sources, addressing challenges in both short-term and long-term epidemic forecasting. This approach allows EINNs to learn latent dynamics while efficiently handling heterogeneous data, improving the adaptability and accuracy of predictions. Millevoi \textit{et al.}~\cite{millevoi2024physics} employ a reduced-split approach using PINNs, which tracks temporal changes in parameters like transmission rates and state variables, achieving enhanced forecasting accuracy and computational efficiency compared to traditional joint training methods. Hu \textit{et al.}~\cite{hu2022modified} apply PINNs to the Susceptible-Infected-Confirmed-Recovered-Deceased (SICRD) model, focusing on estimating unknown infected compartments and parameters by introducing wavelet transforms for data preprocessing and modifying the loss function to handle multiple unknowns, significantly improving prediction reliability. Tang \textit{et al.}~\cite{tang2023enhancing} introduce the Multi-scale Spatial Disease prediction Network (MSDNet), which combines macroscopic population flow data with microscopic contact patterns, effectively capturing disease spread dynamics while incorporating mobility factors like travel restrictions. This hybrid approach leads to a 15-30\% improvement in prediction accuracy over existing models. These studies highlight the fusion of physical models with AI techniques, enabling more accurate, adaptable, and efficient epidemic forecasting in urban environments, while offering flexible tools that can adapt to the dynamic nature of disease spread. By combining physical laws with machine learning, these efforts set a new standard for epidemic modeling, providing critical insights for real-time public health interventions.

\paragraph{Safety and social security} Recent studies highlight the effectiveness of combining physics theories with AI techniques to tackle complex urban safety challenges. Zhang \textit{et al.}~\cite{zhang2022physics} introduce a PIML framework that combines traditional physics-based models with neural networks to enhance the accuracy of crowd simulation and dynamics, offering more reliable predictions for urban safety. Li \textit{et al.}~\cite{li2024physics} apply PINNs to predict crowd density by incorporating fluid dynamics principles, significantly improving the efficiency and precision of crowd movement models. Arun \textit{et al.}~\cite{arun2023physics} integrate safety field theory, inspired by electromagnetic fields, with AI techniques to estimate crash risk and severity, providing a more context-sensitive and accurate approach to traffic safety. Lee \textit{et al.}~\cite{lee2024physics} leverage a data-driven PINN framework to predict vehicle and pedestrian trajectories, using Monte Carlo simulations to assess potential collisions and enhance preemptive risk mitigation. These studies demonstrate how combining physical laws with machine learning enhances model accuracy and generalizability, offering more effective urban safety management strategies. The fusion of AI with physical models allows for improved handling of dynamic interactions between pedestrians, vehicles, and infrastructure, addressing limitations of traditional methods. The integration of data-driven insights with physical modeling provides robust tools for real-time risk evaluation and mitigation. Furthermore, these innovations showcase the potential of hybrid approaches to enhance predictive capabilities across diverse urban scenarios. Ultimately, the application of PIAI offers transformative solutions for crowd safety, traffic management, and broader urban social security.

\paragraph{Service supply and planning}

Recent research in urban service supply and planning highlights the integration of physics theories with AI techniques to address complex urban challenges. Jenkins \textit{et al.}~\cite{jenkins2023physics} fuse high-resolution physics-based hazard simulations with AI to predict the impact of climate change and multi-hazard events, providing critical insights for resilient urban planning. Mao \textit{et al.}~\cite{mao2021urban} develop the Urban Weather Generator (UWG), which combines physics-based microclimate simulations with AI to optimize energy efficiency and sustainability in urban buildings. Su \textit{et al.}~\cite{su2024large} propose a PIAI approach to solving the Facility Location Problem by using graph theory and reinforcement learning to optimize facility placement and maximize accessibility with reduced computational costs. These studies demonstrate how the fusion of physical models, like transportation flow, climate interactions, and hazard simulations, with AI methods can provide adaptive, scalable solutions for urban planning. The integration of physics-based knowledge with AI enhances the predictive accuracy and efficiency of urban service allocation, improving the overall resilience of cities. By leveraging domain-specific physical laws, these models offer valuable decision-making tools for future-proofing urban infrastructures and promoting sustainable urban development.

\subsection{Emergency Management}
\begin{figure}
    \centering
    \includegraphics[width=\linewidth]{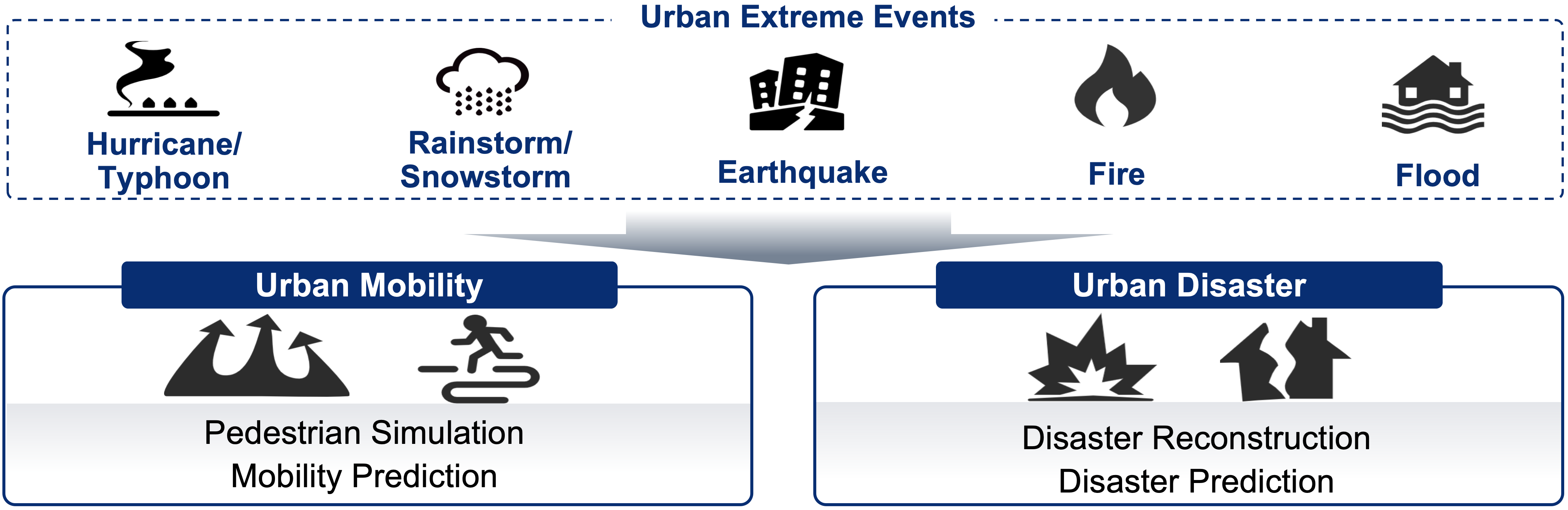}
    \caption{The applications of PIAI in Emergency Management.}
    \label{fig:4.7}
\end{figure}

\subsubsection{Research Focus and Challenges in Urban Emergency Management.}
As we discuss in Section~\ref{sec:physicallaws}, researchers in the domain of urban emergency management mainly focus on planning, coordinating, and executing strategies to respond to and mitigate disasters like natural calamities, industrial accidents, or other public safety threats, which is referred to as urban extreme events in the following paper.
To achieve this goal, numerous methods are developed to sense and predict disaster dynamics and human behavior under the influence of urban extreme events.
Therefore, in the following part of this section, we will first introduce the useful physical theories in urban emergency management. Then, we will discuss existing works leveraging PIAI to reconstruct, predict, or simulate human behavior and disaster dynamics.

\subsubsection{Physical Theories in Urban Emergency Management}
Within the domain of urban emergency management, numerous physical theories are utilized to model urban dynamics and human dynamics. In terms of human behavior, a well-known physical theory is the social force model~\cite{helbing1995social,liu2024tdnetgen}, which is able to model the microscopic human movement under the influence of urban extreme events and has been discussed detailedly in Section~\ref{sec:physicallaws}. Another famous physical theory describing macroscopic human movement under the influence of urban extreme events is the spatio-temporal decay model~\cite{li2022spatiotemporal}, which can be represented as
$
r_i(t) = {\overline{r_i}}/{(1+k(t){\textstyle \sum_{j=1}^{L}w_{ij}N_j(t))}},
$,
where $\overline{r_i}$ is the activity of human mobility without the influence of urban extreme events, $r_i(t)$ is the activity of human mobility under the influence of urban extreme events, and $N_j(t)$ is the metric of the spatiotemporal disaster intensity for geographical region $j$.
Furthermore, to predict disaster dynamics, diverse physical theories have been systematically applied to model the distinct physical processes underlying various extreme events. For instance, fluid dynamics principles are employed for flood prediction, while WRF–SFIRE are employed for wildfire forecasting.
We will discuss them in detail in the following sections.

\paragraph{Mobility Simulation and Prediction for Emergency Management}
Simulating and Predicting human mobility is crucial for emergency management, as it can effectively support a range of emergency response tasks.
In terms of microscopic mobility behavior, pedestrian simulation is instrumental for avoiding incidents like crowd pushing during disasters.
In terms of macroscopic mobility behavior, individual mobility or mobility flow prediction are important for applications including location-based early disaster warnings, pre-distribution of rescue resources, and planning for humanitarian aid, etc.
Specifically, Li et al.~\cite{li2024physics} utilize the physical knowledge from the spatiotemporal decay model~\cite{li2022spatiotemporal} in the way of Method 6 to predict the post-disaster population flow.
Zhang et al.~\cite{zhang2022physics} and Chen et al.~\cite{chen2024social} utilize the physical knowledge from the social force model to simulate crowd dynamics, where physical knowledge are integrated into neural networks in the manner of Methods 3 and 6, respectively. These methods enable a better understanding and prediction of human behavior from micro-to-macro scales, thereby reducing the impact of disasters on urban residents.

\paragraph{Disaster Modeling and Prediction for Emergency Management.}
At the same time, a large portion of the work still focuses on leveraging PIAI methods to better modeling and predicting disaster, thereby enabling more effective urban emergency management.
\citeauthor{lee2024predicting}~\cite{lee2024predicting} utilize the data generated by the physical model of hydrodynamics to train their AI models to predict peak inundation depths of floods, respectively.
\citeauthor{tong2023earthquake}~\cite{tong2023earthquake} draw inspiration from the delay-embedding theorem to detect earthquake precursors from geodetic data.
\citeauthor{borate2023using}~\cite{borate2023using} incorporate fault zone physics into predictive deep learning models utilizing PINNs.
Tang et al.~\cite{tang2025predicting} draw inspiration from the multi-scale characteristics of cascade failures for interdependent infrastructure networks to predict them effectively.
\citeauthor{zhang2023skilful}~\cite{zhang2023skilful} integrate the advective conservation and convective initiation dynamics into a neural network architecture for extreme precipitation nowcasting.
\citeauthor{dabrowski2023bayesian}~\cite{dabrowski2023bayesian} solve the level-set equation in the way of PINN for for wildfire fire-front modeling.
\citeauthor{bottero2020physics}~\cite{bottero2020physics} utilize PIAI methods for better wildfire propagation prediction by reimplementing key components of the WRF-SFIRE wildfire simulator by replacing traditional numerical solvers.
\citeauthor{sitte2022velocity}~\cite{sitte2022velocity} and \citeauthor{shaddy2024generative}~\cite{shaddy2024generative} utilize PINNs, to reconstruct velocity fields in puffing pool fires and infer wildfire arrival times from satellite data, respectively. Overall, these methods advance the precision and efficiency of disaster modeling across diverse hazards by combining well-established physical knowledge and powerful deep learning models.

\subsection{Urban System as a Whole}

Urban systems are inherently interdependent and dynamic, comprising multiple coupled subsystems. Traditional modeling approaches have largely focused on individual subsystems in isolation, which limits their ability to capture cross-domain interactions and systemic behaviors. Recent research highlights the importance of holistic modeling that integrates these diverse domains, providing a more faithful representation of urban complexity. PIAI offers a promising paradigm for such integrated modeling, as it combines the explanatory power of physical laws with the flexibility of data-driven learning, enabling the joint analysis of heterogeneous urban processes \cite{mylonas2021digital,karniadakis2021physics}.

\paragraph{Cross-Domain Dynamics and Multi-Scale Integration}
The evolution of PIAI for holistic urban modeling has progressed from early efforts targeting individual subsystems to increasingly sophisticated frameworks capable of capturing cross-domain dynamics and multi-scale interactions. Early studies primarily focused on embedding physical constraints into single-domain models, such as traffic flow prediction \cite{huang2020physics} or building energy modeling \cite{chen2023physics}. With advances in PIAI methodologies, researchers have shifted toward integrated modeling of interacting urban subsystems. Lin et al.\ developed a physics-informed graph neural network that couples flood dynamics with traffic modeling, embedding hydrodynamic diffusion equations into a GNN architecture to improve traffic flow prediction during urban flooding events \cite{lin2024enhancing}. Li et al.\ proposed a physics-informed neural ODE model for post-disaster human mobility recovery, incorporating physical movement constraints to better simulate population flow dynamics under disrupted conditions \cite{li2024physics}. Tang et al.\ introduced an integrated graph-based framework for modeling cascading failures across interdependent infrastructure networks—including power, transportation, communication, and buildings—achieving superior accuracy in forecasting systemic disruptions \cite{tang2025predicting}.

Recent works further extend PIAI applications to routine urban operations. For example, Rethnam and Thomas combined urban microclimate models with building thermal physics to predict city-wide indoor comfort levels \cite{ramalingam2024physics}, while de Giuli et al.\ applied physics-informed neural networks for modeling and optimization of district heating systems, providing efficient and physically consistent surrogate models for coupled thermal energy networks \cite{de2024physics}. These developments illustrate the expanding scope of PIAI from isolated subsystem modeling to unified treatment of heterogeneous dynamics across domains. PIAI not only supports multi-scale integration—bridging local dynamics (e.g., building-level thermal behavior) with city-wide or regional trends (e.g., energy flows or urban climate patterns)—but also enables multi-source data fusion while maintaining physical consistency \cite{ramalingam2024physics,de2024physics}. With these capabilities, PIAI is increasingly regarded as a powerful tool for urban system-of-systems modeling, providing enhanced predictive accuracy, interpretability, and generalization for complex urban scenarios \cite{tang2025predicting,ramalingam2024physics}.

\section{Discussions and Open problems}
\subsection{Selecting Physical-AI Fusion Methods for Specific Tasks}
Selecting physical-AI fusion methods in urban computing requires understanding the core strengths and limitations of each paradigm. Physics-based models, grounded in conservation laws, offer strong interpretability and robustness in data-scarce scenarios \cite{barthelemy2019statistical}, but struggle with nonlinear complexity and high-dimensional efficiency. AI models, by contrast, excel in capturing complex patterns and enabling fast inference \cite{jin2023spatio}, yet rely heavily on data and often lack physical consistency.

Method selection should consider three key factors: the degree of physical knowledge, data availability, and spatiotemporal scale. Combined with task-specific goals—such as accuracy, efficiency, and interpretability—these dimensions form a decision space for model design. Physics-dominant models suit tasks with strong physical constraints; deep learning with physical validation fits weaker constraints \cite{ribeiro2023mathematical}. Hybrid models guided by physical priors are ideal for sparse data, while data-rich settings favor AI models with embedded physical regularization \cite{silva2019urban}. Real-time tasks benefit from lightweight physics-informed networks; offline tasks allow more complex, tightly coupled architectures.
Beyond task-specific strategies, a unified, general-purpose physical-AI foundation model is a promising direction. Current approaches remain largely data-driven; early explorations like UniST and UrbanDiT \cite{yuan2024unist,yuan2024urbandit} introduce knowledge-guided prompts, yet further efforts are needed to embed physical priors and enhance generalization across diverse urban contexts.

\subsection{Limitations and Future Directions of PIAI in Urban Systems}
Physics-AI fusion methods face several challenges in urban computing \cite{yuan2023spatio}. PINNs lack theoretical guarantees for convergence and generalization, limiting their reliability in high-stakes scenarios. Their training is resource-intensive, hindering real-time deployment \cite{shi2021physics}. Existing models also struggle with multi-scale phenomena like turbulence and chaos, and though physics priors reduce data demands, high-quality data is still needed for calibration. Beyond formal physical laws, urban systems involve rich relational knowledge, such as causality and domain rules—often best represented through knowledge graphs \cite{liu2023urbankg,liu2022developing}. Yet current frameworks rarely incorporate such structured, non-symbolic knowledge, limiting interpretability and reasoning. Scalability remains a bottleneck, constraining applications in large-scale digital twins and real-time simulations.

Future research should strengthen theoretical foundations, improve training efficiency, and explore architectures like multi-resolution networks to handle urban complexity better \cite{ji2022stden}. Data-efficient strategies—e.g., active and transfer learning—are also essential. Cross-disciplinary collaboration will be key to integrating physics-AI approaches into real-world urban systems.
A promising direction is combining PINNs with large language models to enhance expressiveness and enable multimodal reasoning \cite{yan2024opencity,feng2024citybench,xu2023urban}. Hybrid architectures and co-training strategies are needed to fully exploit this potential. Looking ahead, developing world models—simulation frameworks that learn, predict, and reason about urban dynamics—offers a path toward interpretable, adaptive, and generalizable digital twins \cite{ding2024understanding}. By fusing physical laws, relational knowledge, and generative learning, world models may serve as the next-generation engines for urban intelligence.

\subsection{Key Findings on PIAI in Urban Systems}
PIAI methods offer great potential for urban systems, enabled by the structural complexity of cities and the coexistence of physical laws and abundant data. Urban infrastructure encompasses diverse, well-defined physical mechanisms across its core subsystems, while modern sensing technologies provide high-frequency, heterogeneous, and multimodal data. This combination of strong physical priors and rich observational input forms a robust foundation for deploying PIAI models in practical urban scenarios.
From a modeling perspective, data-driven strategies often favor AI-dominant approaches to enhance predictive accuracy, especially when dealing with large-scale and noisy datasets. These methods excel in capturing nonlinear dynamics and extracting high-dimensional patterns. In contrast, urban governance and decision-making applications tend to prefer physically grounded models due to their interpretability and alignment with known laws \cite{pollock2016policy}. The dual nature of urban systems—marked by both human-driven uncertainty and physical regularity, along with complex interdependencies across subsystems—demands a balance between physical interpretability and AI’s representational power. As a result, urban computing emerges as a natural arena for interdisciplinary innovation, where PIAI methods can address critical challenges and drive progress across science and practice.

\subsection{Open problems}
Despite the promising progress of PIAI in urban systems, several critical open problems remain. First, the theoretical foundation of PIAI methods, particularly regarding convergence, generalization, and error bounds in complex, high-dimensional urban environments, remains underdeveloped. Second, there is a pressing need for unified frameworks that can seamlessly integrate heterogeneous physical laws—ranging from deterministic equations to stochastic and agent-based models—into AI architectures for comprehensive urban modeling. Third, current methods lack robustness when faced with noisy, incomplete, or biased urban data, highlighting the necessity for advanced data assimilation, uncertainty quantification, and domain adaptation techniques. Fourth, most PIAI applications remain domain-specific; developing transferable and generalizable models that can adapt across different urban subsystems is an open challenge. Fifth, the integration of symbolic reasoning, causal inference, and structured knowledge representations into PIAI remains limited but crucial for enhancing interpretability and decision support. Lastly, practical deployment of PIAI in real-world urban digital twins is still constrained by scalability, computational cost, and the lack of systematic validation against dynamic, multi-source urban data streams. Addressing these open problems will be pivotal for advancing the scientific and practical impact of PIAI in complex urban systems.
\section{Conclusion}
This survey provides a comprehensive synthesis of PIAI methods in the context of complex urban systems. By proposing a structured classification spanning AI-dominated, AI-physics balanced, and physics-dominated approaches, and detailing seven representative methodologies, this work offers a clear framework for understanding and selecting PIAI strategies tailored to different urban applications. The integration of physical knowledge with data-driven techniques has shown remarkable potential in enhancing predictive accuracy, interpretability, and robustness across key urban domains such as energy, environment, transportation, economy, information, public services, and emergency management. Leveraging both the rigor of physical laws and the flexibility of AI, PIAI enables more reliable modeling of dynamic and multi-scale urban processes, thereby providing powerful tools for improving urban planning, decision-making, and system resilience.


\bibliographystyle{ACM-Reference-Format-num}
\bibliography{ref}

\appendix

\counterwithout{table}{section} 
\renewcommand{\thetable}{S\arabic{table}} 

\section{Related survey}
\begin{table}[ht]
    \centering
    \caption{Comparison with existing surveys. This paper provides a comprehensive review of PIAI methods and their applications in urban systems.}
    \label{tab:surveys}
    \resizebox{\textwidth}{!}{
    \begin{tabular}{llll}
    \hline
    \textbf{Survey} & \textbf{Venue and Year} & \textbf{Main Focus} & \textbf{Deficiency} \\
    \hline
    \cite{carleo2019machine}  & RMP, 2019         & Physical problems            & Limited to physics-based analysis \\
    \cite{cuomo2022scientific}  & JSC, 2022       & Technical methods            & Not targeted at urban systems \\
    \cite{hao2022physics}  & Arxiv, 2022          & Technical methods   & Not targeted at urban systems \\
    \cite{karniadakis2021physics}  & Nat. Rev. Phys., 2021         & Physical problems           & Not targeted at urban systems \\
    \cite{lei2020data}   & IEEE TPWRS, 2020     & Power flow            & Limited scope \\
    \cite{xu2023physics}   & RESS, 2023         & Reliability and system safety      & Limited scope \\
    \cite{wu2024physics}   & ESWA, 2024         & Anomaly and condition monitoring      & Limited scope \\
    \hline
    \end{tabular}
    }
\end{table}

\section{Summary of Physical Laws and Models in Urban Systems}
\begin{table}[ht]
    \centering
    \caption{Typical physical laws in different urban systems.}
    \resizebox{\linewidth}{!}{
    \begin{tabular}{cccc}
    \hline
    \textbf{Domain}                          & \textbf{Element}      & \textbf{Physical model}              & \textbf{Equation}                                                                                                                                                                                                                                                                                                                                              \\ \hline
    \multirow{2}{*}{\textbf{Energy}}         & Electricity           & Power Flow Calculation~\cite{smith2022effect}               &   $\frac{d^2\theta_i}{dt^2}+\gamma\frac{d\theta_i}{dt}=P_i-\kappa\sum_{j=1}^{n}A_{ij}sin\left(\theta_i-\theta_j\right)$                                                                                                                                                                                                                                    \\ \cline{2-4} 
                                             & Gas                   & Gas Network Pressure Balancing Model~\cite{zlotnik2015optimal} & $\sum_{j\in\mathcal{N}\left(i\right)}C_{ij}\cdot\mathrm{sgn}\left(p_i-p_j\right)\cdot\sqrt{\left|p_i^2-p_j^2\right|}=d_i$                                                                                                                                                                                                                                      \\ \hline
    \multirow{3}{*}{\textbf{Environment}}    & Water                 & Storm Water Management Model~\cite{chang2015novel}         & $\frac{\partial h}{\partial t}=\frac{\sum_{k} Q_{p_k}+\sum_{k}   Q_{i_k}-Q_s}{A_m}$                                                                                                                                                                                                                                                                            \\ \cline{2-4} 
                                             & Air                   & Gaussian plume model~\cite{arystanbekova2004application}                 & $C\left(x,y,z\right)=\frac{Q}{2\pi\sigma_y\sigma_zu}\exp{\left(-\frac{\left(y-y_0\right)^2}{2\sigma_y^2}\right)}\exp{\left(-\frac{\left(z-z_h\right)^2}{2\sigma_z^2}\right)}$                                                                                                                                                                                  \\ \cline{2-4} 
                                             & Sound                 & Sound Propagation Model~\cite{bies2023engineering}              & $L_p=L_w-20\log_{10}{\left(r\right)}-11$                                                                                                                                                                                                                                                                                                                       \\ \hline
    \multirow{2}{*}{\textbf{Economy}}        & Goods                 & Network Flow Model~\cite{ahuja1993network}                    & $\sum_{j\in\mathcal{N}^-\left(i\right)}   f_{ji}-\sum_{k\in\mathcal{N}^+\left(i\right)} f_{ik}=b_i$                                                                                                                                                                                                                                                            \\ \cline{2-4} 
                                             & Consumer              & Huff model~\cite{huff1963probabilistic}                           & $P_{ij}=\frac{A_j\cdot d_{ij}^{-\beta}}{\sum_{k\in\mathcal{N}}   A_k\cdot d_{ik}^{-\beta}}$                                                                                                                                                                                                                                                                    \\ \hline
    \multirow{3}{*}{\textbf{Transportation}} & Railway               & OpenTrack~\cite{nash2004railroad}                            & $F_{\mathrm{resistive}}=C_r\cdot m\cdot   g+C_d\cdot\frac{1}{2}\cdot\rho\cdot A\cdot v^2$                                                                                                                                                                                                                                                                      \\ \cline{2-4} 
                                             & Road                  & SUMO~\cite{behrisch2011sumo}                                 & \begin{math}\begin{aligned} \frac{dv}{dt}=&a\left(v,\rho\right)=a_{\mathrm{max}}\left(1-\left(\frac{v}{v_{\mathrm{max}}}\right)^\beta\right)\\&-\gamma\cdot\rho\cdot\left(v-v_{\mathrm{leader}}\left(t\right)\right)\end{aligned}\end{math}                                                                                                                    \\ \cline{2-4} 
                                             & Vehicle               & Car Following and Lane Change Model~\cite{treiber2000congested}  & $\Delta x=\frac{1}{\rho}-\frac{v}{\beta}$                                                                                                                                                                                                                                                                                                                      \\ \hline
    \multirow{2}{*}{\textbf{Information}}    & Communication Network & Communication Network Model~\cite{shannon1948mathematical}          & $C=B\log_2{\left(1+\frac{S}{N}\right)}$                                                                                                                                                                                                                                                                                                                        \\ \cline{2-4} 
                                             & Base Station          & Ray Tracing Model~\cite{rappaport2024wireless}                    & $L\left(d\right)=L_0+10n\log_{10}{\left(\frac{d}{d_0}\right)}+X_\sigma$                                                                                                                                                                                                                                                                                        \\ \hline
    \textbf{Public Services}                 & Public health         & SIR Model~\cite{kermack1927contribution}                            & \begin{math}\begin{aligned} \frac{dS}{dt} &= -\beta \frac{S I}{N} \\ \frac{dI}{dt} &= \beta \frac{S I}{N} - \gamma I \\ \frac{dR}{dt} &= \gamma I \end{aligned}\end{math}                                                                                                                                                                                      \\ \hline
    \textbf{Emergency Management}            & Pedestrian            & Social Force Model~\cite{helbing1995social}                   & \begin{math}\begin{aligned} m \cdot \frac{d^2 \mathbf{r}_i}{dt^2} =& -k_{\text{goal}}   (\mathbf{r}_i - \mathbf{r}_{\text{goal}}) + \sum_{j \neq i}   \frac{\gamma}{|\mathbf{r}_i - \mathbf{r}_j|^2} \hat{\mathbf{r}}_{ij} \\ & +  \sum_{k \in \text{obstacles}} \frac{\beta}{|\mathbf{r}_i - \mathbf{r}_k|^2}   \hat{\mathbf{r}}_{ik} \end{aligned}\end{math} \\ \hline
    \end{tabular}
    }
    \label{tbl:physical_laws}
\end{table}

\section{Summary of Representative Papers for Urban Subsystems}
\begin{table}[ht]
    \centering
    \caption{Summary of representative papers on energy systems.}
    \label{tab:energy}
    \resizebox{\textwidth}{!}{
    
    \begin{tabular}{lllll}
    \hline
    Domain                              & Paper                         & Physical model                                                 & Physical Theory type                     & PINN Methods \\ \hline
    \multirow{3}{*}{Power generation}   & \cite{pombo2022assessing}     & Power generation principle of solar and wind power. & Physical models                          & Method 4    \\
                                        & \cite{zhang2021three}         & 3-D Navier–Stokes equations.                                   & Physical equations                       & Method 1 \& 6 \\
                                        & \cite{liu2024physics}         & Power converter designing constraints.  & Physical models                          & Method 1 \& 3. \\ \hline
    \multirow{4}{*}{Power transmission} & \cite{fioretto2020predicting} & Electrical circults theory.                                    & Physical equations                       & Method 1    \\
                                        & \cite{nellikkath2022physics}  & Electrical circults theory.                                    & Physical equations                       & Method 1 \& 5 \\
                                        & \cite{zhang2024physics}       & Simulation of distributed grids.                               & Expert knowledge & Method 2    \\
                                        & \cite{li2023federated}        & Operation rules of distributed grids.                          & Rule-based constraints                   & Method 1    \\ \hline
    \multirow{3}{*}{Power consumption}  & \cite{qu2024physics}          & Relationship between electricity price and demand.             & Rule-based constraints                   & Method 4    \\
                                        & \cite{gokhale2022physics}     & 2R2C model in building thermal modeling.                       & Physical models                          & Method 1 \& 7 \\
                                        & \cite{ma2024pump}             & Navier-Stokes equations in pipes.                              & Physical equations                       & Method 1    \\ \hline
    \end{tabular}
    }
\end{table}

\begin{table}[ht]
	\centering
	\caption{Summary of representative papers on environment systems.}
	\label{tab:environment}
	\resizebox{\textwidth}{!}{
		\begin{tabular}{lllll}
			\toprule
			\textbf{Domain} & \textbf{Paper} & \textbf{Physical Model} & \textbf{Physical Theory Type} & \textbf{Fusion Method}\\
			\midrule
			\multirow{7}{*}{Air} 
                & \cite{airphynet_iclr24} & Diffusion-Advection PDE & Physical equations & Method 6 \\
                & \cite{tian2024air} & Boundary-Aware Diffusion-Advection & Physical equations & Method 5 \\
                & \cite{mei2024traffic} & Finite Volume and PDE Models & Physical equations & Method 6 \\
                & \cite{bukhari2022fractional} & Fractional Lorenz Chaos Model & Physical models & Method 4 \\
                & \cite{li2023improving} & Pollutant Advection-Diffusion PDE & Physical equations & Method 1 \\
                & \cite{wei2023indoor} & Navier-Stokes and Continuity & Rule-based constraints; Physical equations & Method 1 \\
                & \cite{anwar2024integrating} & Mobility Dynamics & Physical models & Method 7 \\
                \hline
                \multirow{9}{*}{Water} 
                & \cite{adombi2024causal} & HBV Groundwater Dynamics & Physical models & Method 4 \\
                & \cite{bandai2022forward} & Richardson-Richards Hydrodynamics & Physical equations & Method 1 \& 6 \\
                & \cite{liang2019physics} & Surface Runoff Advection-Diffusion & Physical models; Data & Method 4 \\
                & \cite{ashraf2024physics} & Flow Conservation and Pressure Relations & Physical laws & Method 4 \\
                & \cite{bertels2023physics} & Mass Dynamics in Shallow Water Equations & Physical equations & Method 3 \\
                & \cite{zhang2022physics} & Hydro-Mechanical PDEs & Physical equations & Method 5 \\
                & \cite{bihlo2022physics} & Nonlinear Shallow Water Dynamics & Physical models & Method 1 \\
                & \cite{dazzi2024physics} & Enhanced Shallow Water Models & Rule-based constraints & Method 1 \\
                & \cite{lan2024reconstructing} & Richards Hydrodynamics & Physical models & Method 5 \\
                \hline
                \multirow{7}{*}{Soil} 
                & \cite{lu2022deep} & Kinematic, $CO_2$, and Heat Equations & Physical equations & Method 6 \\
                & \cite{ouyang2024machine} & Solid Waste System Dynamics & Physical models & Method 6 \\
                & \cite{oikawa2024inverse} & Battery Degradation Models & Physical models & Method 6 \\
                & \cite{zhang2022physics} & 2D Soil Consolidation PDE & Physical models & Method 6 \\
                & \cite{singh2024piml} & Pile-Soil Dynamics Models & Physical models & Method 1 \\
                & \cite{seydi2024predictive} & Richards Soil Hydrodynamics & Physical equations & Method 5 \\
                & \cite{xie2024simulating} & Terzaghi Consolidation PDE & Physical models & Method 1 \& 6 \\
                \hline
                \multirow{2}{*}{Waste} 
                & \cite{he2023novel} & Radar Backscatter Dynamics & Expert knowledge & Method 2 \\
                & \cite{tao2025non} & Vegetation-Soil Combustion Physics & Physical equations & Method 4 \\
                \hline
                \multirow{2}{*}{Carbon} 
                & \cite{nathaniel2023above} & Ecosystem Productivity Models & Physical models & Method 3 \\
                & \cite{bedi2025neural} & Chemical Transport Model & Physical equations & Method 5 \\
			\bottomrule
		\end{tabular}
	}
\end{table}

\begin{table}[ht]
	\centering
	\caption{Summary of representative papers on transportation systems.}
	\label{tab:transportation}
	\resizebox{\textwidth}{!}{
		\begin{tabular}{lllll}
			\toprule
			\textbf{Research Domain} & \textbf{Paper} & \textbf{Physical Model} & \textbf{Physical Theory Type} & \textbf{Fusion Method}\\
			\midrule
			\multirow{8}{*}{State Estimation} 
                & \cite{huang2020physics, huang2022physics} & LWR \& CTM & Physical models & Method 1 \& 4 \\
                & \cite{zhao2022observer, zhao2024observer} & ARZ & Physical models & Method 1 \\
                & \cite{shi2021physics, shi2021fundamental, shi2021hybrid} & LWR \& ARZ  & Physical models & Method 1 \\
                & \cite{zhao2023learning, huang2024incorporating} & non-local LWR model & Physical models & Method 1 \\
                & \cite{ka2024physics} & Generalized bathtub model & Physical models & Method 1 \\
                & \cite{mo2022quantifying, mo2022traffic} & LWR \& ARZ & Physical models, Uncertainty & Method 1 \& 4 \\
                & \cite{wang2024knowledgedatafusionorientedtraffic} & $\alpha$-SPIDL \& $\beta$-SPIDL & Physical models & Method 1 \& 3 \\
                & \cite{shi2023spatiotemporal} & LWR & Physical models, Expert knowledge & Method 1 \& 2 \\
                \hline
                \multirow{2}{*}{Data Imputation} 
                & \cite{tang2024physics} & CTM & Physical models & Method 1 \\
                & \cite{xue2024network} & $\lambda$-trapezoidal MFD & Physical models & Method 1 \\
                \hline
                \multirow{4}{*}{Flow Prediction}
                & \cite{yuan2022trafficflowmodelingwithgradual} & LWR, ARZ \& CTM & Physical models & Method 1 \\     
                & \cite{ji2022stden} & PEF, Differential Equation & Physical equations & Method 1 \\
                & \cite{li2022network} & MFD & Physical models & Method 4 \& 1 \\
                & \cite{deshpande2024kalman} & LWR & Physical models & Method 7 \\
                \hline
                \multirow{2}{*}{Control} 
                & \cite{su2021adaptive} & Kinematic wave model & Physical models & Method 3 \& 4 \\
                & \cite{han2022reinforcement} & PI-ALINEA & Control strategies & Method 2 \\
			\bottomrule
		\end{tabular}
	}
\end{table}

\begin{table}[ht]
	\centering
	\caption{Summary of representative papers on economy systems.}
	\label{tab:economy}
	\resizebox{\textwidth}{!}{
		\begin{tabular}{lllll}
			\toprule
			\textbf{Research Domain} & \textbf{Paper} & \textbf{Physical Model} & \textbf{Physical Theory Type} & \textbf{Fusion Method}\\
			\midrule
			\multirow{7}{*}{Manufacturing} 
                & \cite{zobeiry2021physics} & Heat Transfer Equations & Physical equations & Method 1 \& 2 \\
                & \cite{fu2024physics} & Steel Frame Mechanics and Code Constraints & Rule-based constraints, Physical laws & Method 5 \\
                & \cite{raabe2023accelerating} & Thermodynamics and Microstructure Evolution & Expert knowledge & Method 1 \& 2 \\
                & \cite{dutta2024application} & Hottel’s Furnace Radiation Model & Physical models & Method 1 \\
                & \cite{wang2022hybrid} & Physical Knowledge in Simulations & Physical laws & Method 1 \& 3 \\
                & \cite{shi2024physics} & Structural Reliability Insights & Physical models & Method 1 \& 6 \\
                & \cite{huang2022physics} & Industrial Load Physics & Physical laws & Method 1 \& 3 \\
                \hline
                \multirow{2}{*}{Stock} 
                & \cite{gatta2023meshless} & Option Pricing Models & Physical models & Method 1 \& 5 \\
                & \cite{aboussalah2024ai} & Option Pricing Core Physics & Physical models & Method 1 \& 4 \\
                \hline
                \multirow{1}{*}{Commodity} & \cite{tao2024predicting} & Corporate Growth Dynamics & Physical models & Method 5 \\
                \hline
                \multirow{2}{*}{POI} 
                & \cite{yang2024siamese} & User Check-In Modeling & Physical equations & Method 7 \\
                & \cite{zhuang2024tau} & Trajectory Flow Transition Model & Physical models & Method 4 \\
			\bottomrule
		\end{tabular}
	}
\end{table}

\begin{table}[ht]
	\centering
	\caption{Summary of representative papers on information systems.}
	\label{tab:information_system}
	\resizebox{\textwidth}{!}{
		\begin{tabular}{lllll}
			\toprule
			\textbf{Research Domain} & \textbf{Paper} & \textbf{Physical Model} & \textbf{Physical Theory Type} & \textbf{Fusion Method}\\
			\midrule
			\multirow{4}{*}{Information diffusion} 
                & \cite{tang2024msa} & User Activity Level Equations & Expert knowledge & Method 7 \\
                & \cite{tu2022modeling} & Heat Transfer Equation & Physical laws & Method 6 \\
                & \cite{cheng2024information} & Probabilistic Diffusion Equations & Physical Equations & Method 6 \\
                & \cite{yu2024information} & Probabilistic Diffusion Model & Physical Equations & Method 1\\
                \hline
                \multirow{4}{*}{Opinion Evolution Modeling} 
                & \cite{de2016learning} & Jump-Diffusion Stochastic Differential Equations & Physical Equations & Method 6 \\
                & \cite{okawa2022predicting} & Hegselmann-Krause Model & Physical models & Method 6 \\
                & \cite{li2024hides} & Neural ODEs and Higher Derivative Constraint & Physical Equations & Method 6 \\
                & \cite{bi2024duan} & Bidirectional Graph ODEs & Physical models & Method 5 \\
			\bottomrule
		\end{tabular}
	}
\end{table}

\begin{table}[ht]
	\centering
	\caption{Summary of representative papers on public service systems.}
	\label{tab:physics_for_public service}
	\resizebox{\textwidth}{!}{
		\begin{tabular}{lllll}
			\toprule
			\textbf{Domain} & \textbf{Paper} & \textbf{Physical Model} & \textbf{Physical Theory Type} & \textbf{Fusion Method}\\
			\midrule
			\multirow{5}{*}{Public Health} 
                & \cite{kharazmi2021identifiability} & Fractional-order epidemiological models & Physical models & Method 6 \\
                & \cite{rodriguez2023einns} & SEIRM and SIRS model & Physical models & Method 1 \\
                & \cite{millevoi2024physics} & Compartmental epidemiological models & Physical models & Method 2 \\
                & \cite{hu2022modified} & SICRD compartmental model & Physical models & Method 3 \\
                & \cite{tang2023enhancing} & Multi-scale SIS model & Physical equations & Method 3 \\
                \hline
                \multirow{4}{*}{Safety and social security} 
                & \cite{zhang2022physics} & Social force model & Physical equations & Method 7 \\
                & \cite{li2024physics} & Conservation of mass principle & Physical laws & Method 4 \\
                & \cite{arun2023physics} & Road user safety field model & Physical equations & Method 4 \\
                & \cite{lee2024physics} & Intelligent driver model; Social force model & Physical equations & Method 3 \\
                \hline
                \multirow{3}{*}{Service supply and planning} 
                & \cite{jenkins2023physics} & Natural hazard models & Physical models & Method 4 \\
                & \cite{mao2021urban} & Four urban weather models & Physical models & Method 4 \\
                & \cite{su2024large} & Facility location knowledge & Expert knowledge & Method 2 \\
			\bottomrule
		\end{tabular}
	}
\end{table}

\begin{table}[ht]
	\centering
	\caption{Summary of representative papers on emergency management systems.}
	\label{tab:information_system}
	\resizebox{\textwidth}{!}{
		\begin{tabular}{lllll}
			\toprule
			\textbf{Research Domain} & \textbf{Paper} & \textbf{Physical Model} & \textbf{Physical Theory Type} & \textbf{Fusion Method}\\
			\midrule
			\multirow{3}{*}{\makecell[l]{Mobility Simulation\\and Prediction}}
                & \cite{li2024physics} & Spatio-temporal Decay Model & Physical Equations & Method 7 \\
                & \cite{zhang2022physics} & Social Force & Physical Equations & Method 4 \\
                & \cite{chen2024social}  & Social Force & Physical Equations & Method 7 \\
                \hline

                \multirow{11}{*}{\makecell[l]{Disaster Modeling\\and Prediction}}
                & \cite{lee2024predicting} & HEC-RAS 2D model & Physical Models & Method 1 \\ 
                & \cite{tong2023earthquake} & Delay-embedding Theorem & Physical Models & Method 3 \\ 
                & \cite{borate2023using} & Fault elastic coupling and stiffness model & Physical Constraints & Method 1 \\ 
                & \cite{tang2025predicting} & Multi-scale representation & Expert Knowledge  & Method 2 \\ 
                & \cite{okazaki2022physics} & Dislocation model & Physical Equations  & Method 1 \\ 
                & \cite{zhang2023skilful}  & \makecell[l]{Advective conservation\\and convective initiation dynamics} & Physical Models  & Method 7\\ 
                & \cite{dabrowski2023bayesian} & Level-set equation & Physical Equations & Method 1 \\  
                & \cite{bottero2020physics} & WRF-SFIRE wildfire simulator & Physical Models & Method 7\\  
                & \cite{sitte2022velocity}  &Reacting Navier–Stokes equations & Physical Equations & Method 1\\  
                & \cite{shaddy2024generative} & WRF–SFIRE  wildfire simulator & Physical Models &  Method 1\\ 
			\bottomrule
		\end{tabular}
	}
\end{table}

\end{document}